\documentclass[11pt,a4paper]{article} 
\usepackage{jheppub}

\usepackage[export]{adjustbox}
\usepackage{float}
\usepackage{caption}
\usepackage{subcaption}
\usepackage{epsfig,multicol,bbm}
\usepackage{graphicx}
\usepackage{amsmath}
\usepackage{slashed}
\usepackage{bm}
\usepackage{multirow}
\allowdisplaybreaks[1]

\newcommand{\ET}{\mbox{$\not \hspace{-0.10cm} E_T$ }}
\newcommand\fverb{\setbox\fverbbox=\hbox\bgroup\verb}
\newcommand\fverbdo{\egroup\medskip\noindent%
			\fbox{\unhbox\fverbbox}\ }
\newcommand\fverbit{\egroup\item[\fbox{\unhbox\fverbbox}]}
\newbox\fverbbox

\title{Simplified DM models with the full SM gauge symmetry : the case of $t$-channel colored scalar 
 mediators}

\author[a,b]{P. Ko,}
\author[a]{Alexander Natale,}
\author[c]{Myeonghun Park,}
\author[b]{and Hiroshi Yokoya}

\affiliation[a]{School of Physics, KIAS, Seoul 02455, Korea}
\affiliation[b]{Quantum Universe Center, KIAS, Seoul 02455, Korea}\affiliation[c]{Center for Theoretical Physics of the Universe, IBS,  Daejeon 34051, Korea}

\emailAdd{pko@kias.re.kr}
\emailAdd{alexnatale@kias.re.kr}
\emailAdd{hyokoya@kias.re.kr}
\emailAdd{parc.ctpu@gmail.com}

\arxivnumber{}	
\abstract{
The general strategy for dark matter (DM) searches at colliders currently
relies on simplified models. In this paper, we propose a new  $t$-channel 
UV-complete simplified model that improves the existing simplified DM models 
in two important respects: (i) we impose the full SM gauge symmetry including the fact
that the left-handed and the right-handed fermions have two independent mediators 
with two independent couplings, and (ii) we include the renormalization group 
evolution when we derive the effective Lagrangian for DM-nucleon scattering 
from the underlying UV complete models by integrating out the $t$-channel mediators. 
The first improvement will introduce a few more new parameters compared with
the existing simplified DM models.  In this study we look at the effect this broader set 
of free parameters has on direct detection 
and the mono-$X$ + MET ($X$=jet,$W,Z$) signatures at 13 TeV LHC while
maintaining gauge invariance of the simplified model under the
full SM gauge group.  We find that
the direct detection constraints require DM masses less than 10 GeV in
order to produce phenomenologically interesting collider signatures.  
Additionally, for a fixed mono-W cross section it is possible to see
very large differences in the mono-jet cross section when the usual simplified
model assumptions are loosened and isospin violation between RH and
LH DM-SM quark couplings are allowed.

}
\begin{document} 
\begin{flushright}
CTPU-16-12 % IBS pre-print no.
\end{flushright}
\maketitle

%---------------------------------------------------------------------------
\section{Introduction\label{sec:intro}}
%---------------------------------------------------------------------------
The astrophysical evidence for the existence of Dark Matter (DM) is convincing, but the
properties of DM remain largely unknown~\cite{DMrev}.  In an effort to elucidate the properties of DM
there are several predominant strategies: indirect detection experiments which search for DM annihilation
signals, direct detection via nuclear recoil experiments such as LUX, and search strategies at colliders where 
DM is directly produced and observed via large missing transverse momentum (\ET \hspace{-2mm}).
When investigating direct detection signals  the effective field theory (EFT) approach is a sensible
way to describe the interaction of DM with the detector while utilizing only two free parameters; 
the scale of the new physics that mediates this DM-SM interaction ($\Lambda_i$ which is 
much larger than the hadronic energy scale) and the DM mass ($m_{\chi}$).   
 The lowest dimensional 
effective Lagrangian for DM direct detection (DD) can be written schematically as 
\begin{equation}
{\cal L}_{\rm DD} = \sum_i \frac{1}{\Lambda_i^2}  ~\bar{q} \Gamma_i q ~\bar{\chi} \Gamma_i \chi \  ( + H.c.) .
\end{equation}
Colliders searches for mono $X +$ \ET signatures (with $X = W,g,\gamma,Z$) at the LHC 
have also used this EFT approach during Run-I~\cite{AtlasJetMet8TeV,AtlasLepMet8TeV,AtlasZMet8TeV,CMSLepMet8TeV,AtlasWZMet8TeV}, 
but have also utilized UV-complete models such as Supersymmetry~\cite{DMatColliders}. 

However, at the center-of-mass energies at the LHC this EFT approach would break down~\cite{BeyondEFTDM,BusoniEtAl2014},
which warrants the use of UV-complete models at the expense of introducing many more free parameters.  An alternative approach is the utilization of so-called simplified DM models 
\cite{SimpleDMmet}.  These simplified DM models generically satisfy a few criteria
~\cite{SimplifiedDMNewPhys,SimplifiedDMLHC}:
the simplified models should involve a particle stable enough so it may produces a 
large \ET signature,   the simplified model should respect the unbroken SM gauge group 
at minimum and it should not violate  approximate and global symmetries  of the SM, 
with the ultimate goal of describing interesting collider phenomenology involving \ET
while keeping the number of free parameters to a minimum.  Then the above effective Lagrangian for DM DD is modified as 
\begin{equation}
\frac{1}{\Lambda_i^2}  ~\bar{q} \Gamma_i q ~\bar{\chi} \Gamma_i \chi  
\rightarrow \frac{g_q g_\chi}{m_\phi^2 - s} ~\bar{q} \Gamma_i q ~\bar{\chi} \Gamma_i \chi 
\end{equation}
when we consider the $s$-channel UV completion for $q\bar{q} \rightarrow \phi \rightarrow \chi \bar{\chi}$.
%This presumes that the underlying UV complete models are described by the following 
%Lagrangian:

However this strategy  with simplified DM models have ample room for improvement in two important 
respects.  First of all, the simplified models do not respect the full SM gauge invariance, 
which may be problematic   when they are adopted to DM search studies at high energy colliders.  
At the LHC CM energy, one has  to respect the full SM gauge symmetry, and not just the unbroken subgroup 
of it.    Recently, importance of the full SM invariance, unitarity and gauge invariance with respect to the 
mediators was noticed in a few independent studies~\cite{Kahlhoefer:2015bea,Baek:2015lna,BellEtAl2015}, which will be detailed in the subsequent discussions. 
When we impose the full SM gauge symmetry, we have to realize that the SM fermions have two independent
chiralities, left-handed (LH) and right-handed (RH), and SM gauge interactions are chiral as well. 
Therefore the LH quark and the RH quark would couple to two different colored mediators, 
$\widetilde{q}_L$  and $\widetilde{q}_R$ with two independent couplings $\lambda_L$ and $\lambda_R$ 
(see Sec. 2 for the $t$-channel UV complete Lagrangian and more precise definitions of these parameters,
and also Feynman diagrams in Figs. 4,5 and 6 in Sec. 4).  Then the UV completion generically
calls for two independent propagators of $\widetilde{q}_L$  and $\widetilde{q}_R$, instead of a 
single propagator, Eq.~(1.2).  Only the case of $W$ + \ET would involve a single propagator, 
because $W$  couples only to the LH quark and its partner mediator.  This phenomena is due 
to the facts that (i) the SM fermions in 4-dim spacetime have two independent chiralities, 
(ii) the SM gauge theory is chiral, and (iii) the full SM gauge symmetry is imposed on the UV 
completions.   Then the simplified DM models proposed in this paper would not violate gauge 
invariance and unitarity.  Otherwise one could get physically nonsensible  results.   
 
Secondly, there is a technical issue when one derives the effective Lagrangian suitable for 
direct detection of DM.  One can integrate out the mediator at the mediator mass scale, 
obtaining 4-fermion operators.   However the relevant energy scale for  the DM direct detection 
cross section is order of nuclear energy scale,
and one has to include the renormalization effects from the mediator mass scale down to 
the nuclear energy scale \footnote{This is well known from flavor physics ($K,B$ physics, see 
Ref.~\cite{Buras:2009if} for example) as well as top forward-backward asymmetry 
\cite{Jung:2014kxa}.}.  
This procedure was not included properly in the simplified DM models~\cite{Boveia:2016mrp}, and should be performed
before one derives the constraints on the simplified DM models from the DM direct detection data.
This can be included in a straightforward manner using the renormalization-group analysis for the 
DM-nucleon scattering \footnote{Recently, this issue has been pointed out in 
Ref.~\cite{D'Eramo:2016atc} in the context of the DM simplified models with $s$-channel 
vector mediators.}.  RG evolution can not only change the effective 
coupling strengths at different energy scale, but also generate new operators that were not present when the 
mediators were integrated out at the mediator mass scales~\cite{DeramoEtAl}.  Due to this second effect, there would be more than
one effective operator at nucleon mass scale that are relevant to DM DD. 
In other words, It is not realistic at all to assume that DM DD can be described by a single effective operator, 
in sharp contrast to what earlier literatures did assume using the effective Lagrangian for the DM DD.

In this paper, we propose a minimal simplified DM model that respects the full SM gauge symmetry, 
assuming the DM is a Dirac fermion $\chi$ with negative dark  $Z_2$ parity and introducing $t$-channel
colored mediators. 
Imposing the full SM gauge symmetry in the DM-SM interaction Lagrangian is the new and 
the unique aspect of our proposal, and improves the earlier attempts for simplified DM models 
for collider searches and direct detection of DM.  
The number of new parameters in the simplified DM models with the full SM gauge symmetry is 
usually ''four'', one more than the simplified models in the literature~\cite{DMbenchmark2016}. One extra parameter is 
coming from the second mass scale, which often enters in the Lagrangian when we impose the full
SM gauge symmetry.  This feature has not been noticed in earlier literature, and the interpretation of 
DM search at colliders and in direct detection experiments is modified when this is taken into account.  In addition, we include the RG running effects when we derive the effective Lagrangian for DM-nucleon 
scattering and compare with the bounds from LUX and other DM direct detection experiments.

In the construction of simplified DM models with the full SM gauge symmetry, we impose the following  conditions to the model Lagrangian:
\begin{itemize}
\item Invariance under the full SM gauge symmetry : in many DM models, one often imposes the invariance 
of the model only under the unbroken SM gauge symmetry, $H_{\rm SM} \equiv SU(3)_C \times U(1)_{\rm em}$.   
This may be acceptable for studying direct detection of DM, but not for collider signatures of DM. 
However, at high energy colliders,  one has to impose the full SM gauge symmetry, $G_{\rm SM} \equiv 
SU(3)_C \times SU(2)_L \times U(1)_Y$.  
The importance of taking into account $SU(2)_L$ gauge invariance when 
investigating DM signatures at colliders 
 has previously been pointed out in a recent paper~\cite{BellEtAl2015}, where a potential enhancement
to the mono-$W$ signature was previously found when 
considering unequal mediator couplings to up and down quarks~\cite{BT13}, 
but this enhancement was found to result from spurious longitudinal 
W boson contributions~\cite{BellEtAl2015}.  The important point is that the EFT method 
can break down at a scale on order of the VEV, well before the scale $\Lambda$ 
as generally assumed~\cite{BellEtAl2015}.  
%Sometimes it is straightforward to impose the full SM gauge symmetry
%on model Lagrangians,  but quite often it is not the case, as we will demonstrate below. 
\item Issue of dark (gauge) symmetry responsible for absolute stability or longevity of DM particle : from the previous discussion on the $G_{\rm SM}$ vs. $H_{\rm SM}$, it is also clear that the 
model Lagrangian and phenomenology thereof would depend strongly on what dark (gauge) 
symmetry we assume is responsible for the DM stability or longevity.  Since we don't know 
anything about the dark sector at the moment, we will make the simplest working assumption 
that DM  in our model carries $Z_2$-odd parity, whereas all the SM particles are even under 
$Z_2$.

It is natural to assume that DM may have some kind of (its own) gauge symmetries 
\cite{Baek:2013qwa,Baek:2013dwa, Ko:2014nha,Ko:2014bka,Ko:2014eia,Ko:2014loa,Baek:2014kna,Ko:2014lsa,Ko:2015ioa,Ko:2015nma}.  
And then there could be  other extra dark fields (such as dark Higgs or dark gauge fields) 
which might be not that heavy and so we may have to include them in our simplified models. 
This part will be highly model dependent, and we make a simple  assumption they are all heavy 
enough so that we can ignore them in our simplified model.  

\item Renormalizability and unitarity : In EFT approaches to DM, it is common to 
consider higher dimensional nonrenormalizable operators.  This approach is a fine starting point,
especially for the DM direct detection. However one has to think about the UV completions eventually, and 
there could be more than one UV completion that leads to the same low energy EFT at a given order. 
Many DM simplified models start from DM direct detection and then extrapolate to collider 
signatures at higher energies.   However, in a series of papers on Higgs portal DM models, 
it has been shown that the EFT can  give us completely misleading results compared with 
the full renormalizable and unitary DM models, 
in the context of singlet fermion DM~\cite{Baek:2011aa,Baek:2012uj} and vector DM \cite{Baek:2012se,Baek:2013dwa,Ko:2014gha} with Higgs portal interactions.  
Since we can easily miss important phenomenology within EFT which is nonrenormalizable 
and nonunitary (see for example Refs.~\cite{Baek:2014jga,Ko:2016xwd}), we will start 
from renormalizable and unitary DM models 
\footnote{In this paper, we consider the colored scalar mediators in the $t$-channel.
If we consider the vector mediators in the $s$-channel, we have to address the issue of gauge anomaly cancellation, which was discussed in Ref.~\cite{Kahlhoefer:2015bea}. }.  
  
\item Flavor physics : If the mediator carries nontrivial SM gauge charges 
(such as color and/or electric charges),  the one loop diagrams involving the DM and the 
mediators may generate the nontrivial FCNC,  which would be strongly constrained by various 
data from the $K, B$ meson systems.  For the case of Dirac fermion DM, the constraints are 
weaker than the case of the Majorana fermion DM, since there is no chirality flip in the loop. 
   
\end{itemize}
Based on these assumptions, we construct minimal simplified DM models with full SM gauge 
symmetry.  In this paper, we shall concentrate on the $t$-channel colored mediators in $q\bar{q} \rightarrow \chi \overline{\chi}$.

This paper is organized as follows. In Sec. 2, we show the simplified DM models with the full
SM gauge invariance as well as renormalizability and unitarity. In Sec. 3, we derive the effective
Lagrangian relevant for DM direct detection by integrating out the colored scalar mediators and
performing the RG evolution down to the nucleon mass scale, and discuss that there appear 
a number of different operators appears simultanesouly. In particular isospin violation would be
generic because of two independent scalar mediators originating from two different chirality
of the SM fermions. In Sec. 4, we derive the amplitudes for  mono $X$ + \ET with $X=W,g$, 
and present the numerical analysis and the releted phenomenogy in Sec. 5. Then we summarize
in Sec. 6.

\section{$t$-channel UV completion with colored scalar mediators}% partners of the SM fermions}
%\subsection{Model}
Let us consider the $t$-channel UV completion with scalars.  We introduce
3 types of new scalar bosons, $\widetilde{Q}_{L i}$, $\widetilde{u}_{R i}$ and $\widetilde{d}_{R i}$
with negative $Z_2$ parity, which 
are partners of $Q_{L i} \equiv ( u_{L i} , d_{L i} )^T$, $u_{R i}$ and $d_{R i}$ respectively 
\footnote{In this paper, we consider only the scalar partners of the SM quarks. It would be straightforward to 
introduce the scalar partners of the SM leptons.}.
Simplified models with colored scalar mediators that couple
to the quarks have been previously studied 
~\cite{monojet8tev,PVZ14,BT13,squarkdmdirect2,squarkdmdirect1,BellEtAl2012,
DMbenchmark2016}, however these studies have assumed either just an up-like $SU(2)$ 
singlet~\cite{squarkdmdirect2}, a down-like singlet~\cite{PVZ14}, a doublet 
~\cite{squarkdmdirect1,BrennanEtAl2016},
 or a simplified model similar to our proposal however with universal 
couplings to all generations and universal masses for up-like and down-like scalars
~\cite{DMbenchmark2016}.
The gauge invariant interaction Lagrangian between quarks and DM in our model is given by:
\begin{equation}
{\cal L} _{t-channel} = - \left[  \overline{\chi} \widetilde{Q}_{L}^{i \dagger}  \left( \lambda_{Q_L} \right)_i^{~j}  
Q_{L j}  + \overline{\chi}  \widetilde{u}_R^{i \dagger} \left( \lambda_{u_R} \right)_i^{~j}  u_{R j}
+ \overline{\chi}  \widetilde{d}_R^{ i \dagger}  \left( \lambda_{d_R} \right)_i^{~j} d_{R j}  + H.c. \right]
\end{equation}
We also show the Lagrangian for the newly introduced scalar fields:
\begin{eqnarray}
{\cal L}_{\rm scalar} & = & D_\mu \widetilde{Q}_L^{i ^\dagger} D^\mu \widetilde{Q}_{L i} 
-   \widetilde{Q}_L^{i \dagger}   \left[ \left( m_{\widetilde{Q}_L,0}^2 \right)_i^{~j} + 2 \left( \lambda_{Q_L H} 
\right)_i^{~j} H^\dagger H \right]  \widetilde{Q}_{L j}    \nonumber  \\ 
& + & D_\mu \widetilde{u}_R^{i \dagger} D^\mu \widetilde{u}_{R i} 
-  \widetilde{u}_R^{i \dagger}   \left[ \left( m_{\widetilde{u}_R,0}^2 \right)_i^{~j} + 2 \left( \lambda_{u_R H} 
\right)_i^{~j} H^\dagger H \right]  \widetilde{u}_{R j}   \nonumber   \\
& + & D_\mu \widetilde{d}_R^{i \dagger} D^\mu \widetilde{d}_{R i} 
-  \widetilde{d}_R^{i \dagger} \left[ \left( m_{\widetilde{d}_R,0}^2 \right)_i^{~j} + 2 \left( \lambda_{d_R H} 
\right)_i^{~j} H^\dagger H \right]   \widetilde{d}_{R j}   \\
& - & \left[  \widetilde{Q}_L^{i \dagger}  \left( A_u \right)_i^{~j} \widetilde{H} \widetilde{u}_{R j} + 
\widetilde{Q}_L^{i \dagger} \left( A_d \right)_i^{~j}   H \widetilde{d}_{R j}  + H.c \right]   \nonumber  \\
& - &  \lambda_{\widetilde{q}_L} ( \widetilde{Q}_L^\dagger \widetilde{Q}_L )^2 
- 2 \lambda_4 H^\dagger \widetilde{Q}_L \widetilde{Q}_L^\dagger H 
%- \left[ 2 \lambda_5 ( H^\dagger \widetilde{q}_L )^2 + H.c.    \right]
 \nonumber  
\end{eqnarray}
where the covariant derivative contains all the SM gauge fields according to the SM gauge quantum
numbers of the fields upon which $D_\mu$ acts.  
At this level, all the fields are in the interaction eigenstates.

The matrices $m_{\widetilde{Q}_L}^2$, $\lambda_{Q_L H}$, $m_{\widetilde{u}_R}^2$, $\lambda_{u_R H}$, 
$m_{\widetilde{d}_R}^2$ and $\lambda_{d_R H} $ are  Hermitian matrices in flavor space. 
We have suppressed the scalar partners of the SM leptons, for which there could be similar terms.
% for the partners of the SM leptons. 

Scalar quark masses are given by 
\begin{eqnarray}
m_{{\widetilde u}_L}^2  & = & m_{\widetilde{Q}_L,0}^2 + \lambda_{Q_L H} v^2  
\nonumber
\\
m_{{\widetilde d}_L}^2  & = &  m_{\widetilde{Q}_L,0}^2 + \lambda_{Q_L H} v^2 
+ \lambda_4 v^2 = m_{{\widetilde u}_L}^2 + \lambda_4 v^2 
\\
m_{{\widetilde u}_R}^2  & = & m_{\widetilde{u}_R,0}^2 + \lambda_{u_R H} v^2 
\\
m_{{\widetilde d}_R}^2  & = & m_{\widetilde{d}_R,0}^2 + \lambda_{d_R H} v^2 
\nonumber   
\end{eqnarray}
Note that the $\lambda_4$ term induces the mass splitting between $\widetilde{u}_L$ and 
$\widetilde{d}_L$: 
\[
m_{{\widetilde d}_L}^2 - m_{{\widetilde u}_L}^2 = \lambda_4 v^2,
\]  
thereby generating isospin violation effects at colliders and at DM direct detections. 
The trilinear $A_{u,d}$ terms generate the left-right mass mixing between $\widetilde{u}_L$ and 
$\widetilde{u}_R$ (and also between $\widetilde{d}_L$ and $\widetilde{d}_R$).  

After EWSB, we have to rotate both quarks and their scalar partners to the mass eigenstates. 
The resulting Lagrangian will be similar to the above one, except that 
\[
H^\dagger H \rightarrow \frac{v^{2}}{2} ( 1 + \frac{h}{v} )^2  .
\]
Note that the $m^2$ and $\lambda$ would not be simultaneously diagonalizable in general. 
Therefore one would have flavor violation in the Higgs couplings to the scalar partners of $Q_L$, $u_R$ 
and $d_R$, which would lead to rare Higgs decays into 
\[
H \rightarrow \widetilde{q}_i^* \widetilde{q}_j^* \rightarrow ( \bar{q}_i + \chi ) + ( q_j + \bar{\chi} ).
\]
Also the scalar partners of the SM quarks will modify $H \rightarrow g g,  \gamma \gamma, Z \gamma$ 
through loop effects.  The deviations of the Higgs signal strengths from the SM values will depend on the
ratio of the Higgs contribution to the mass of the scalar partners of the SM fermions. 

Basically this case is similar to the MSSM,  except that there is only one species of 
neutral dark Dirac fermion in our case.  Once we include 3 generations of dark scalar partners, 
their mass matrices would not be diagonal in the basis where quark masses are diagonal. 
This misalignment of mass matrices in the flavor space would lead to flavor and CP violation 
induced by dark scalars, similarly to the gluino-mediated FCNC and CP violation in the general MSSM.
One crucial difference exists in this model, since the DM is Dirac fermion and not a Majorana
fermion there is no chirality flip inside the loops, and the usual FCNC constraints become
weaker in our model compared to the MSSM.

\section{Direct detection}

Let us derive the effective Lagrangian describing the direct detection cross section for the 
DM-nucleon scattering.  Note that there are a number of different effective operators generated 
simultaneously if we integrate out the dark scalars, $\widetilde{Q}_L$, $\widetilde{u}_R$ and
$\widetilde{d}_R$.  The Wilson coefficient of the effective operators depend on a number
of parameters, including three different mass scales of dark scalars, and there is no single mass
scale we can associate with a single effective operator, in sharp contrast to the conventional 
wisdom. This is due to the condition that the DM interactions with the SM fermions 
respects the full SM gauge symmetry. Both collider searches and 
direct detection of $\chi$ depend on at least two different and independent mass scales.

The resulting effective Lagrangian for dark matter direct detection is given by 
\begin{eqnarray}
{\cal L}_{\rm DD} & = &  - \frac{\left| ( \lambda_{q_L} )_1^{~1} \right|^2}{m_{\tilde{u}_L}^2} 
 \bar{\chi} u_L \bar{u}_L \chi   -  \frac{\left| ( \lambda_{q_L} )_1^{~1} \right|^2}{m_{\tilde{d}_L}^2}  
 \bar{\chi} d_L \bar{d}_L \chi   
- \frac{\left| ( \lambda_{u_R} )_1^{~1} \right|^2}{m_{\tilde{u}_R}^2} 
\bar{\chi} u_R \bar{u}_R \chi   \nonumber   \\
& - & \frac{\left| ( \lambda_{d_R} )_1^{~1} \right|^2}{m_{\tilde{d}_R}^2} 
 \bar{\chi} d_R \bar{d}_R \chi  
% \nonumber   \\
- \frac{\left| ( \lambda_{q_L} )_2^{~2} \right|^2}{m_{\tilde{s}_L}^2}  \bar{\chi} s_L \bar{s}_L \chi    
- \frac{\left| ( \lambda_{d_R} )_2^{~2} \right|^2}{m_{\tilde{s}_R}^2} 
 \bar{\chi} s_R \bar{s}_R \chi  
\end{eqnarray}

After Fierz transformation, the above Lagrangian is cast into the following form:
\begin{eqnarray}
\label{DD-Fierz}
{\cal L}_{\rm DD} & = &  \frac{  \left| ( \lambda_{q_L} )_1^{~1} \right|^2}{2 m_{\tilde{u}_L}^2} 
\bar{\chi}_R \gamma_\mu \chi_R \bar{u}_L \gamma^\mu u_L  
 + \frac{  \left| ( \lambda_{q_L} )_1^{~1} \right|^2}{2 m_{\tilde{d}_L}^2}  \bar{\chi}_R \gamma_\mu \chi_R 
\bar{d}_L \gamma^\mu d_L    \nonumber  \\ 
& + & \frac{ \left| ( \lambda_{u_R} )_1^{~1} \right|^2}{2 m_{\tilde{u}_R}^2} 
\bar{\chi}_L \gamma_\mu \chi_L \bar{u}_R \gamma^\mu u_R  % \nonumber   \\
+  \frac{\left| ( \lambda_{d_R} )_1^{~1} \right|^2}{2 m_{\tilde{d}_R}^2} 
 \bar{\chi}_L \gamma_\mu \chi_L \bar{d}_R \gamma^\mu d_R   
 \nonumber   \\
& + & \frac{ \left| ( \lambda_{q_L} )_2^{~2} \right|^2}{2 m_{\tilde{s}_L}^2}  
\bar{\chi}_R \gamma_\mu \chi_R \bar{s}_L \gamma^\mu s_L      
+ \frac{\left| ( \lambda_{d_R} )_2^{~2} \right|^2}{2 m_{\tilde{s}_R}^2} 
 \bar{\chi}_L \gamma_\mu \chi_L \bar{s}_R \gamma^\mu s_R   
\end{eqnarray}
Since the strange quark current does not contribute to the nucleon matrix element, we can ignore
the strange quark and concentrate only on the 1st generation quarks. Therefore we shall suppress
the generation indices on the matrices $\lambda$'s, and make abbreviations: $\lambda_q$, 
$\lambda_u$ and $\lambda_d$ from now on for the DM direct detection. 

We have worked in the leading order in QCD, ignoring the $\chi \chi G_{\mu}^a G^{a\mu\nu}$ and 
twist-2 operators that could contribute to the DM direct detection.  It would be straightforward to include
them in the analysis of DM direct detection, which have been calculated explicitly
for Majorana DM~\cite{DreesNojiri93,HisanoEtAlGluon,GondoloScopel2013}, and can be used without 
modification for Dirac DM~\cite{IbarraWild2015}.

Because new physics generates a number of effective operators simultaneously in general,
it is unrealistic to assume that, for example, new physics generates only 
$( \bar{\chi} \gamma_\mu \chi ) ( \bar{q} \gamma^\mu q)$  or $( \bar{\chi} \chi ) ( \bar{q} q)$.  
Considering the effective operators suitable for describing DM direct detection  
as starting points, one can miss important pieces of new physics at colliders regarding 
the DM sector.

We have assumed that the couplings $\lambda_{q_L}$, $\lambda_{u_R}$ and 
$\lambda_{d_R}$ are flavor diagonal in the mass eigenstates in order to avoid the bounds 
from the FCNC.  We also assume that the $\widetilde{d}$ and $\widetilde{s}$ are degenerate 
in order to avoid the constraints from $K^0 - \overline{K^0}$ mixing. 
These assumptions are rather ad hoc, but do not violate the underlying gauge symmetry, and
thus are theoretically consistent assumptions.  They could be somewhat relaxed within 
the current constraints from the FCNC in the $K$ and $B$ meson systems compared with the
SUSY case, since we assume that the DM is a Dirac fermion, and not a Majorana fermion, 
unlike the $N=1$ SUSY models.    

Then the direct detection cross sections from DM-N scattering   (with $N=p,n$) are given by 
\begin{eqnarray}
\label{SI}
\sigma^{\rm SI}_N & = & \frac{1}{64 \pi}~\frac{m_N^2 m_\chi^2}{(m_\chi + m_N)^2}~
\left[  \left( \frac{ 3 \left| \lambda_{Q_L} \right|^2}{2 m_{\tilde{Q}_L}^2} 
+ \frac{ \left|  \lambda_{u_R} \right|^2}{2 m_{\tilde{u}_R}^2} 
+ \frac{ 2 \left| \lambda_{d_R} \right|^2}{2 m_{\tilde{d}_R}^2} \right)   \right.
\nonumber  \\
& + & \left.  \frac{Z}{A}  \left(  \frac{ \left| \lambda_{u_R}  \right|^2}{2 m_{\tilde{u}_R}^2} 
- \frac{  \left| \lambda_{d_R} \right|^2}{2 m_{\tilde{d}_R}^2} 
\right)   \right]^2,
\end{eqnarray}
under the assumption that  $\lambda_4=0$.
As discussed numerous places in the literature~\cite{DeramoEtAl,DeramoEtAl2014,HisanoEtAlGluon,HisanoEtAlEW,HisanoEtAl2015}, 
there are potentially important 
direct detection effects due to RGE running from the UV to the nuclear scale.
A user-friendly procedure for approximating these effects is presented in
Ref.~\cite{DeramoEtAl}, and this process
generically induces extra dependencies on the mediator mass in the 
SI cross section and can induce isospin-violating effects and 
different structures of operators than what occurs at the scale of the 
mediator mass.  By implementing the procedure outlined in Ref.~\cite{DeramoEtAl} to account for
the running of the EFT from the scale $\Lambda$ to the hadronic scale, and
taking into account the latest results from LUX~\cite{lux,lux2015},
it is found that for dark matter masses in the range 
10 GeV $< m_{\chi} <$ 1000 GeV, our simplified model is essentially excluded for any choice of $\lambda_{q_i}$ and
$m_{\widetilde{q}_i}$ that would be phenomenologically interesting at colliders.   
Note that the relic density for GeV scale DM is generically
too large for TeV mass scalars \cite{monojet8tev}, and too low when the DM is TeV
scale, however the tension 
with the relic density can potentially be alleviated with
couplings to the leptons and still evade the LUX constraints~\cite{IbarraWild2015}.  However,
the direct detection and relic density constraints were found assuming $\chi$ is absolutely stable, and is the only DM particle.  In scenarios where
$\chi$ is only stable long enough to escape detection at a collider, 
or where there are multiple particles responsible for the total DM energy density in the universe, then the constraints from direct detection 
and the relic density of DM can be relaxed.  Specifically, in the case where
there are two DM species, the direct detection cross section will depend
on the relative density of each DM species divided by its mass times the
relevant cross section \cite{MultipartiteDM}.  Under such assumptions,
it is possible that the $\chi$ which produces the interesting collider signatures
is an order of 100s of GeV while avoiding direct detection and relic density constraints, and in this case the earlier assumption that the mediators which allow coupling to heavier quarks are much larger than the other mediators, allows for the Fermi-LAT constraints on $ b \bar{b}$ annhilation to be satisfied~\cite{fermilat}.  Note that in the case where $m_{\chi}<10$~GeV, the indirect detection constraints are important and would need to be considered in the case where interactions are added, however for the purposes of the paper we do not specify any interactions beyond the $t$-channel mediators.
%For the purposes of this study we focus on GeV scale DM, that is $m_{\chi} < 10 GeV$.  
Additional contributions to the direct detection cross section
can also include contributions from gluons and 
the electric/magnetic dipole moments, however the gluon contribution to direct detection found in Ref.~\cite{HisanoEtAlGluon} is proportional to \( \alpha_s \frac{m_{\chi}}{m_{\widetilde{q}_i}^4} \), 
thus in the low DM mass region ($m_{\chi}<10$ GeV) the gluon operator is suppressed, and the dipole moment contributions 
to direct detection are small when there is a tree-level coupling
to up and down quarks~\cite{IbarraWild2015} as is the case in our model.

\begin{figure}[htb]
\centering
 \begin{minipage}{1\textwidth}
   \centering
  \includegraphics[width=0.90\textwidth]{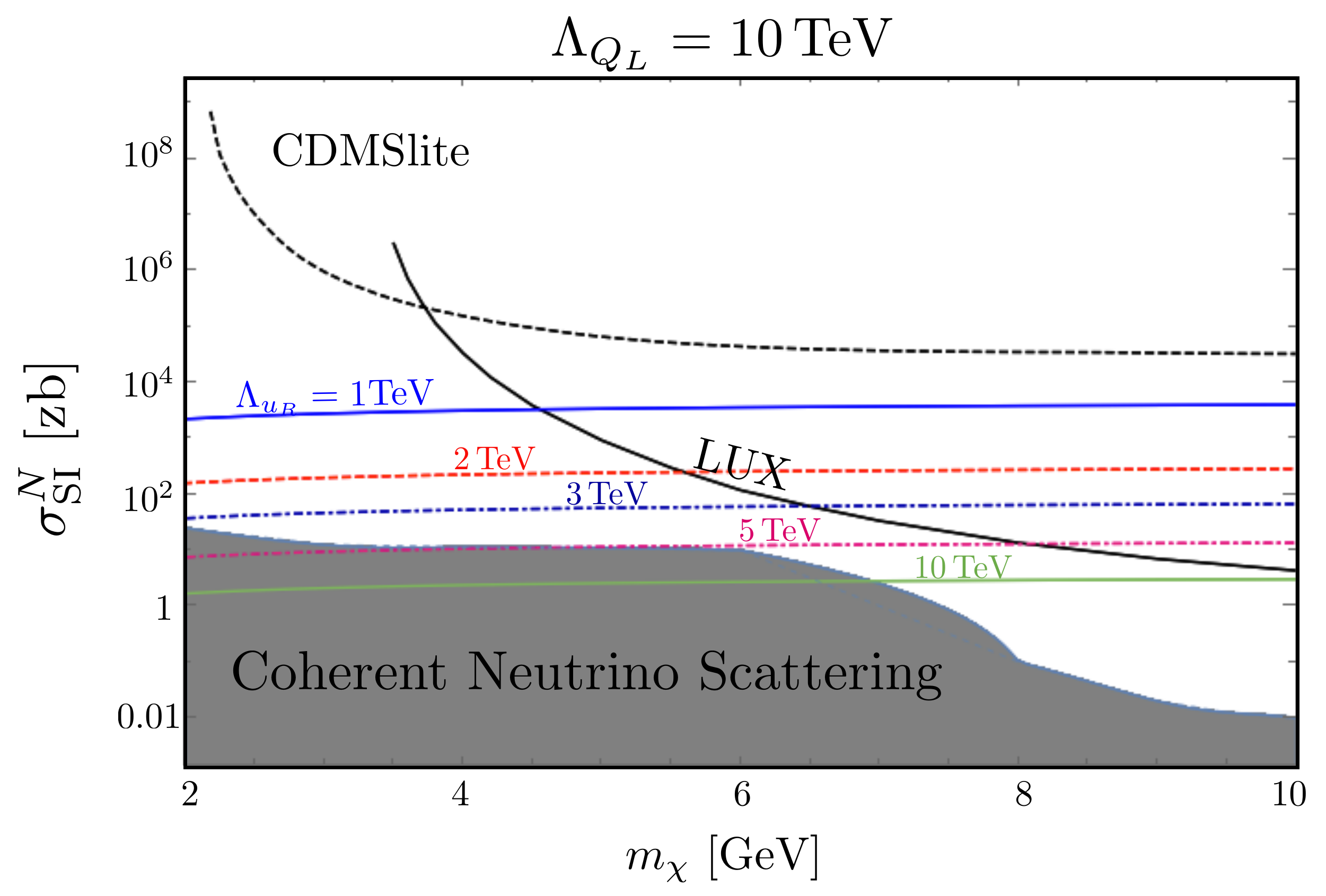}
  \end{minipage}
\caption{\small Low DM mass spin-independent WIMP-Nucleon scattering constraints for $\lambda_{Q_L}=\lambda_{u_R}=1$ and $\lambda_{d_R}=0$.  Effects from 
running are accounted for using the procedure outlined in Ref.~\cite{DeramoEtAl}.}
\label{DDluxCDMS}
\end{figure}

Generic 
isospin violation is also present (even when $\lambda_4=0$)
due to the unbalanced $ \lambda_{u_R} $  and 
$\lambda_{d_R} $, which is consistent 
with the underlying SM gauge symmetry.  Due to the generic isospin violation,
and the mixing of different operators due to running effects noted earlier,
it is not legitimate to consider only one mediator mass,
 and in general we have to allow all three 
different mediators for direct detection (and generally non-zero
values for $\lambda_4$).
That is,
even if a particular set of EFT operators at the EW or TeV scale 
is assumed, the running effects generically produce a
mixture of the EFT operators~\cite{DeramoEtAl2014,DeramoEtAl}.  In Fig.~\ref{DDluxCDMS} the direction detection
cross section, with full RGE running effects taken into account through the procedure outlined in Ref.~\cite{DeramoEtAl}, is shown
versus the $m_{\chi}$.  The dark scalar couplings are assumed to be $\lambda_{Q_L}=\lambda_{u_R}=1$ and $\lambda_{d_R}=\lambda_4=0$.
CDMS Lite Run 1 and Lux direct detection constraints are plotted, along with the coherent neutrino scattering background, which 
severely constrain $m_{\chi}$ if we assume $\chi$ is absolutely stable and the major source of the DM energy density.  It is
important to note that for a fixed $\Lambda_{Q_L}$ variations in $\lambda_{u_R}$
change the direct detection cross-section by almost four orders of magnitude in this low $m_{\chi}$ region.
Even in the simplifying assumption where the doublet is mostly decoupled (i.e., $\Lambda_{Q_L}>10$ TeV)
and $\Lambda_{u_R}=\Lambda_{d_R}=1$ TeV, 
 the SI cross section is close to the limit from Run 1 CDMSlite
~\cite{CDMSlite}.  
Importantly, the generic isospin violating effect leads to a potentially large material dependence, 
for instance using Eq.~\ref{SI}, and the Z and A values 
for Xenon and Germanium, the relative
difference in the cross sections ($ \frac{\sigma_{SI}^{Xe}-\sigma_{SI}^{Ge}}{\sigma_{SI}^{Ge}}$) is given by:
\begin{equation}
\label{DD-rel-diff}
\Delta \sigma / \sigma = \frac{-76 \Lambda^2_{Q_L} (\Lambda^2_{d_R} -\Lambda^2_{u_R}) (1684\Lambda_{d_R}^2
\Lambda_{Q_L}^2 + 7074\Lambda_{d_R}^2 + 1853 \Lambda_{Q_L}^2 \Lambda_{u_R}^2)}{17161 (13 \Lambda_{d_R}^2 \Lambda_{Q_L}^2 + 54 \Lambda_{d_R}^2 \Lambda_{u_R}^2 +14 \Lambda_{Q_L}^2 \Lambda_{u_R}^2)^2},
\end{equation}
\begin{figure}[htb]
%\centering
  \begin{minipage}{0.5\textwidth}
  %\flushleft
   \includegraphics[scale=0.7]{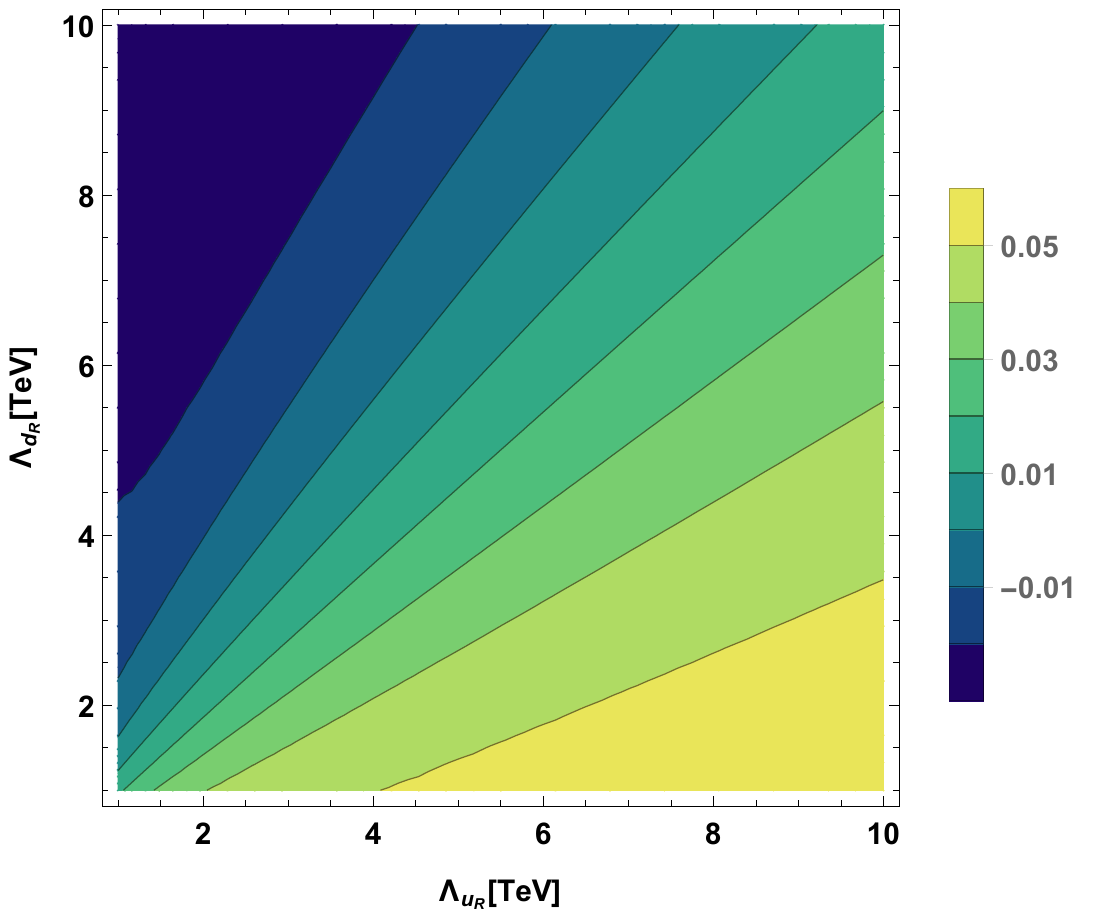}
  \end{minipage}
  \hfill
  \begin{minipage}{0.5\textwidth}
  %\hspace{2em}
  %\flushleft
  \includegraphics[scale=0.7]{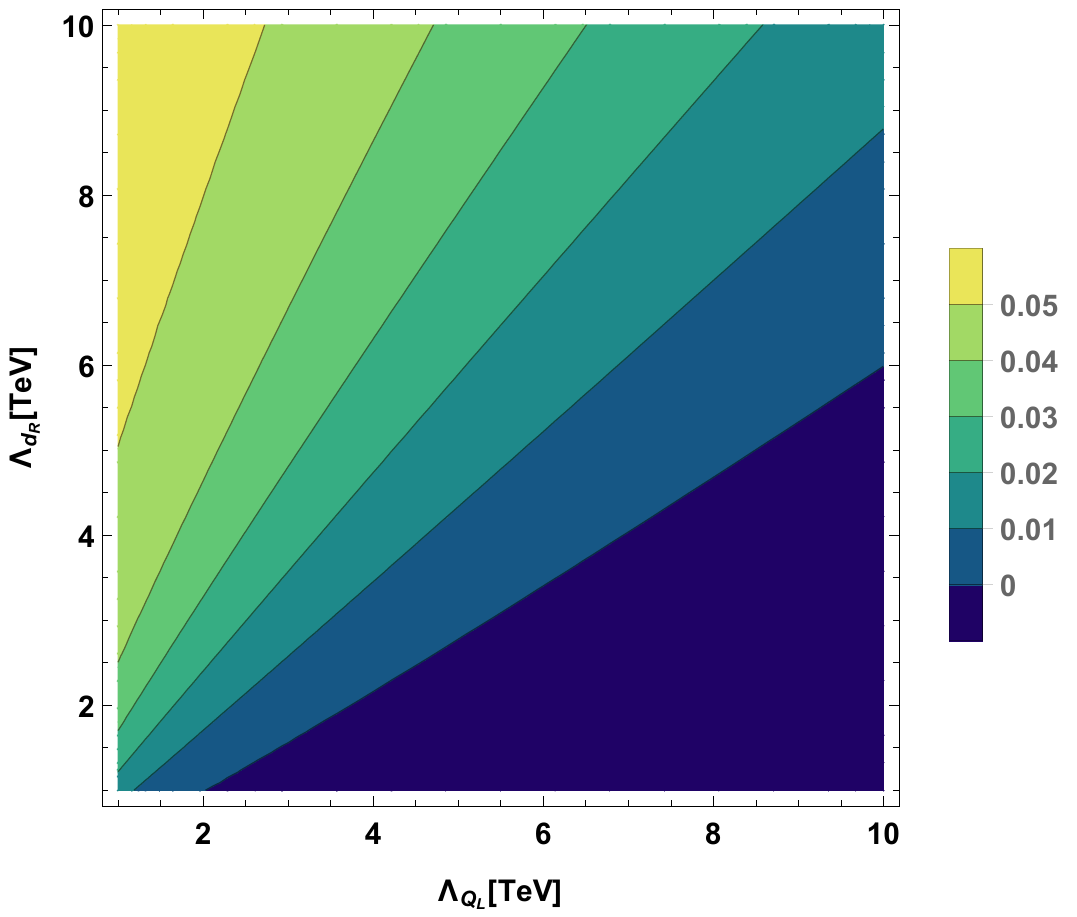}
  \end{minipage}
  \hfill
  
  \centering
  \begin{minipage}{0.5\textwidth}
  \includegraphics[scale=0.7]{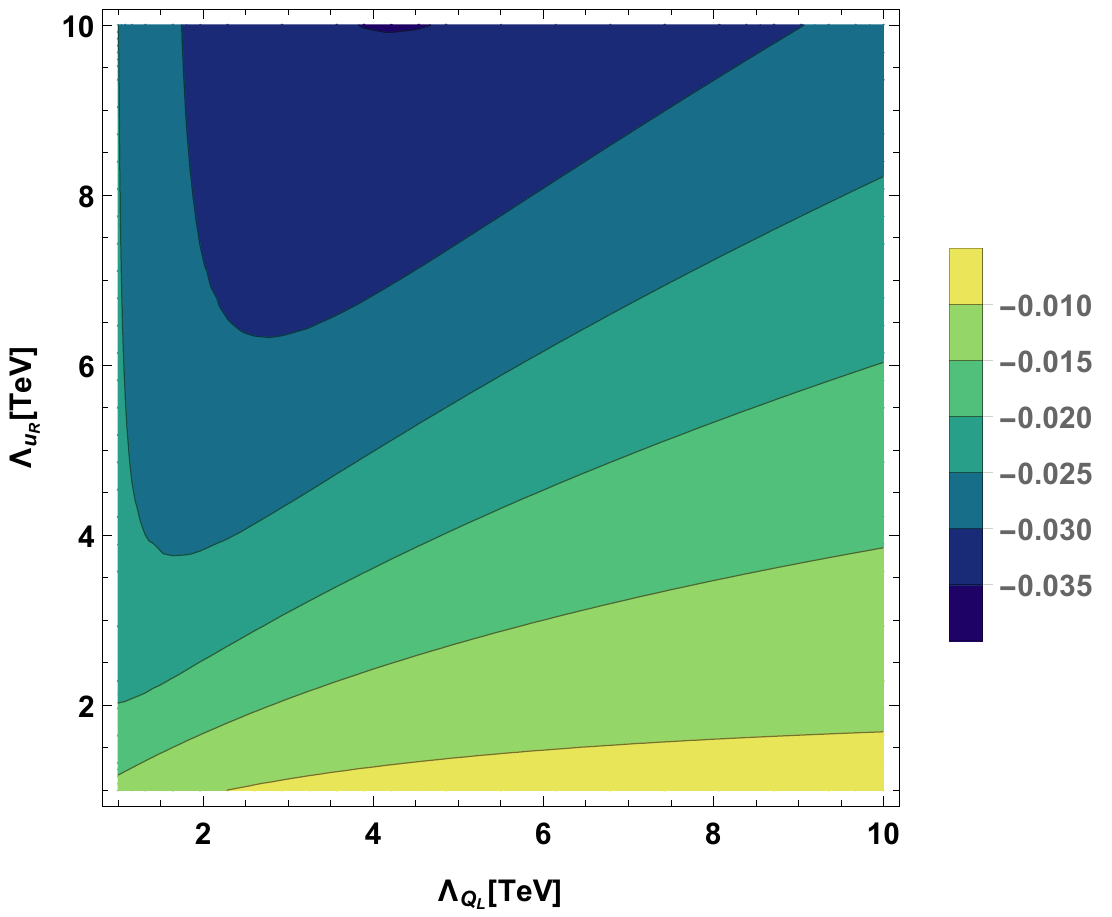}
  \end{minipage}
\caption{$\Delta \sigma/\sigma$ with running effects with one mediator scale entirely decoupled,
and $\lambda_4$.}
\label{DDApprox}
\end{figure}
\begin{figure}[htb]
\centering
  \begin{minipage}{0.45\textwidth}
  %\flushleft
   \includegraphics[scale=0.7]{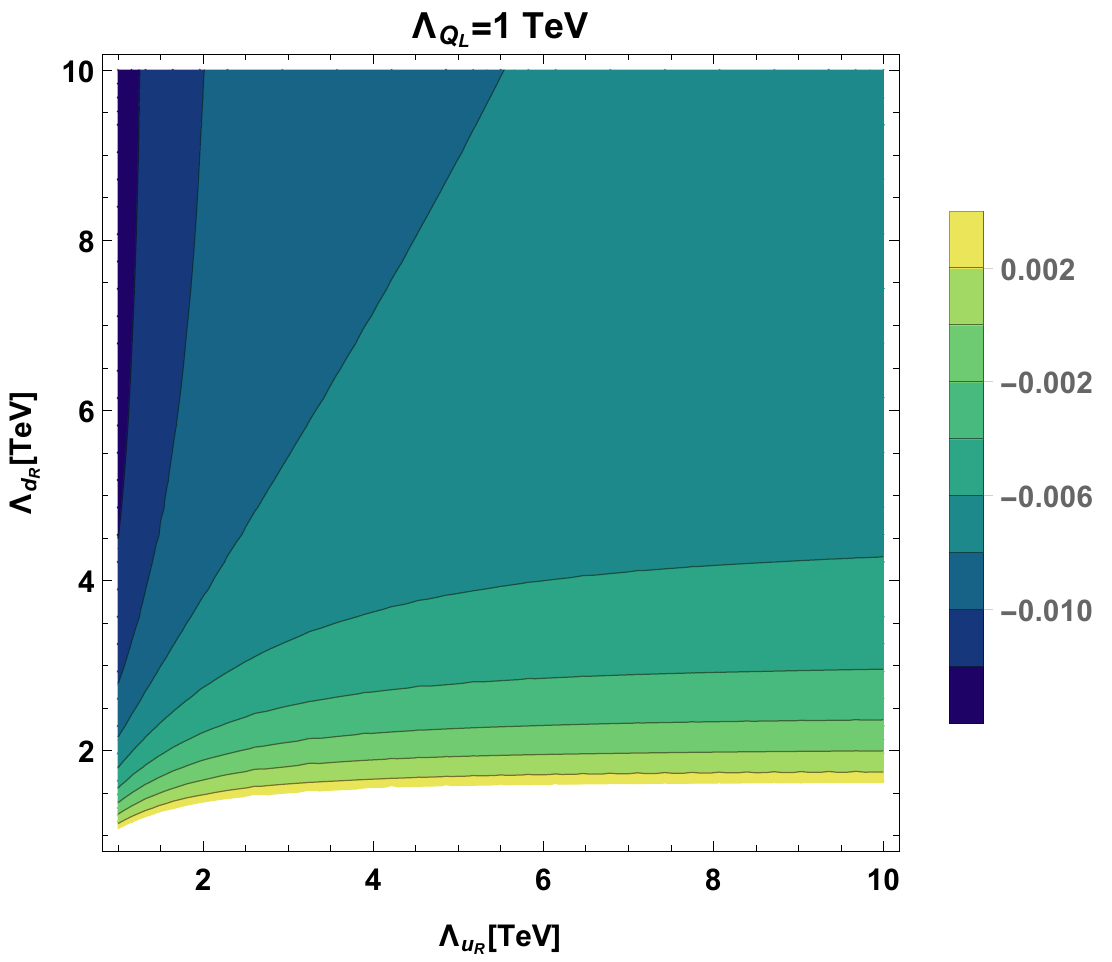}
  \end{minipage}
  \hfill
  \begin{minipage}{0.45\textwidth}
  %\flushleft
  \includegraphics[scale=0.7]{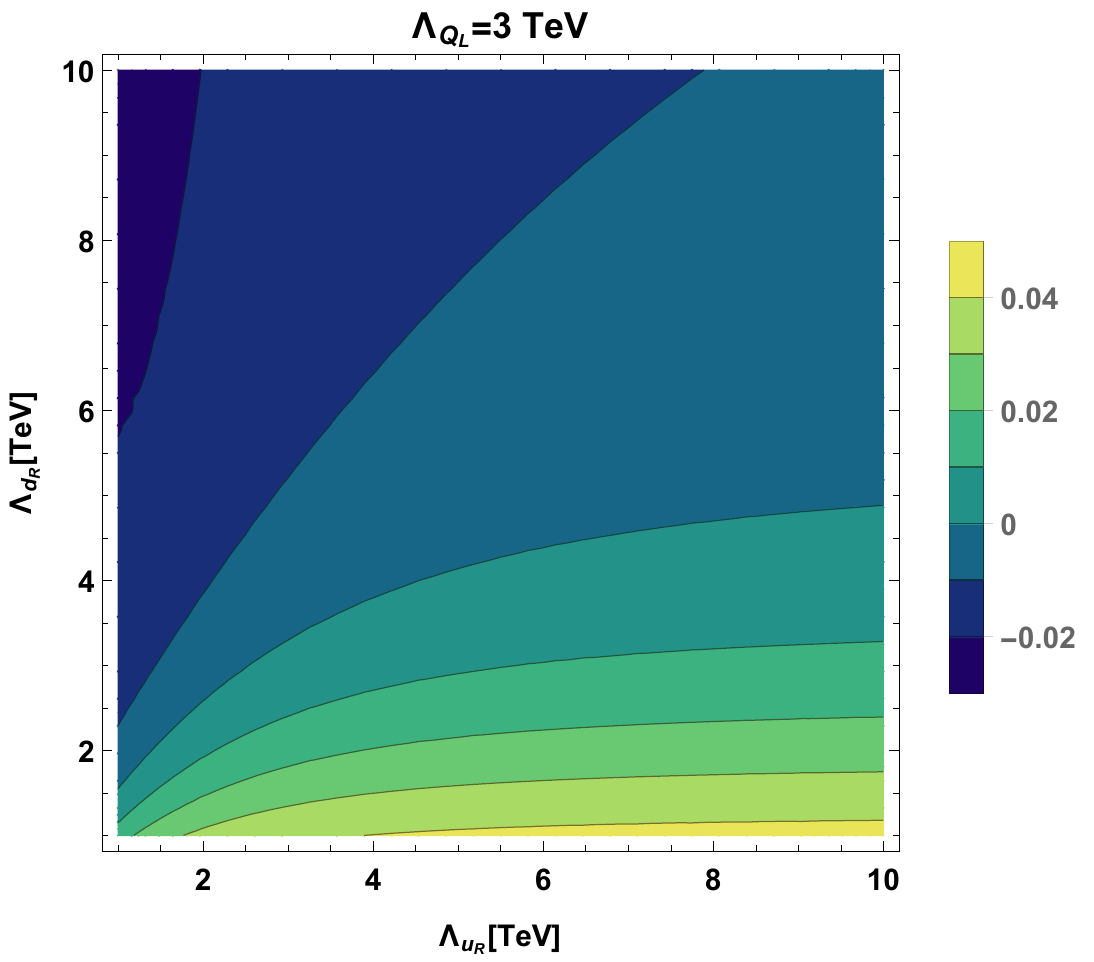}
  \end{minipage}
  \hfill
  \begin{minipage}{0.45\textwidth}
  %\hspace{2em}
  %\flushleft
  \includegraphics[scale=0.7]{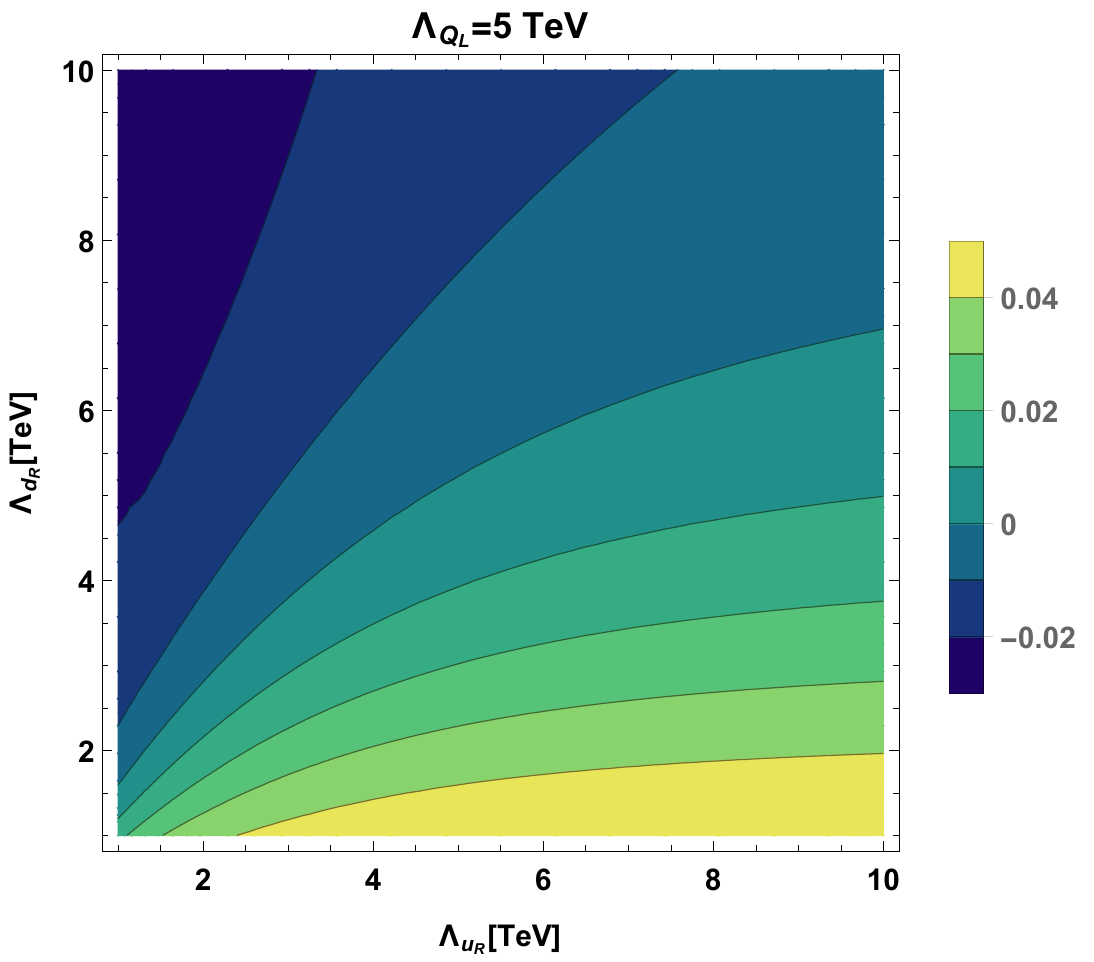}
  \end{minipage}
  \hfill
  \centering
  \begin{minipage}{0.45\textwidth}
  \includegraphics[scale=0.7]{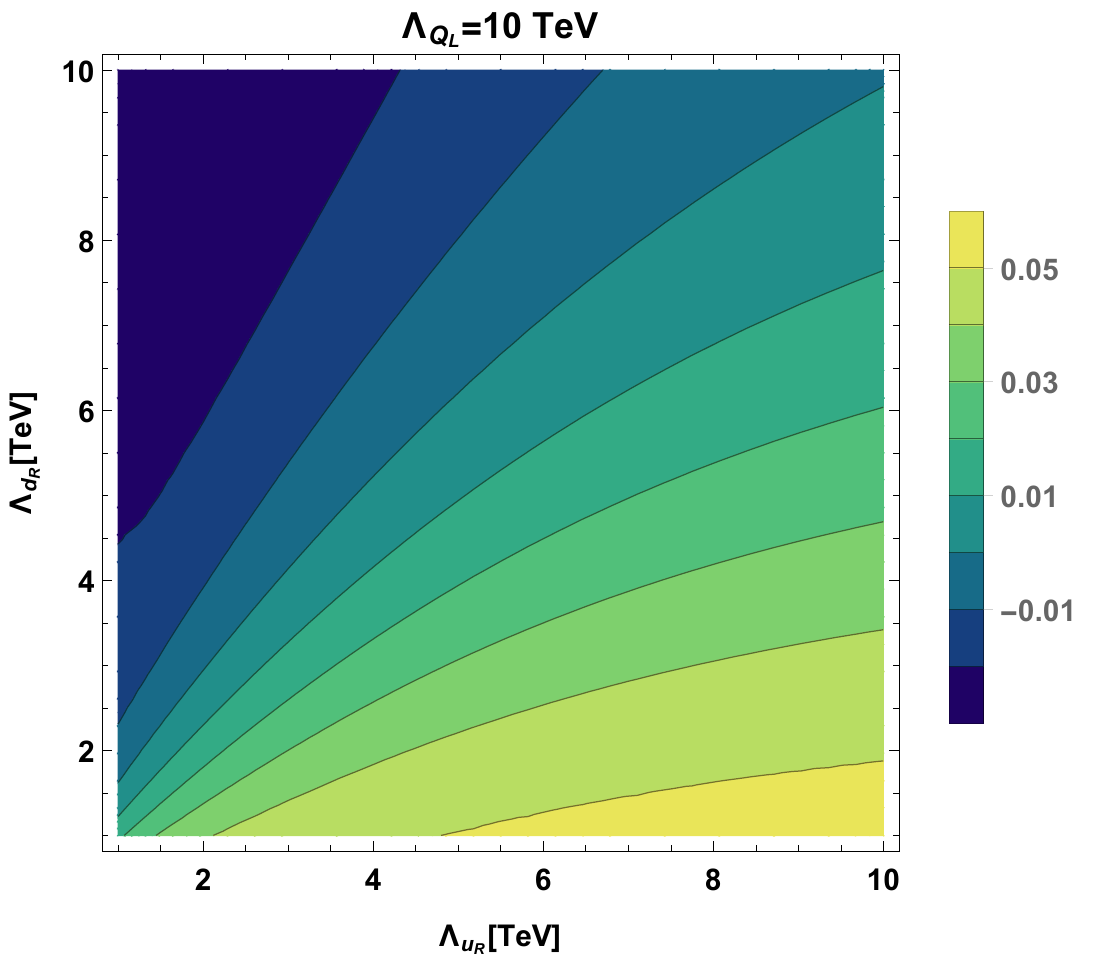}
  \end{minipage}
v\caption{$\Delta \sigma / \sigma$ with running effects for fixed $\Lambda_{Q_L}$ and $\lambda_4=0$.}
\label{DDExact}
\end{figure}
where $\Lambda_{q_i} = m_{\widetilde{q}_i} / \lambda_{q_i}$. This equation becomes zero if $\Lambda_{u_R} = \Lambda_{d_R}$, or if both singlet EFT scales are taken to infinity (i.e., the singlets are entirely decoupled).  However, in the limit where 
$\Lambda_{Q_L},\Lambda_{u_R} \rightarrow \infty$, Eq.~\ref{DD-rel-diff} yields a positive value of $0.0419$ ($\sigma_{SI}^{Xe}>\sigma_{SI}^{Ge}$), where as $\Lambda_{Q_L},\Lambda_{d_R} \rightarrow \infty$ produces a value on order of $-0.0441$ ($\sigma_{SI}^{Xe}<\sigma_{SI}^{Ge}$).  The exact 
expression for the isospin violating effects is modified
from running effects, which can be significant~\cite{DeramoEtAl}.
In Fig.~\ref{DDApprox} a contour plot of $\Delta \sigma / \sigma$ is shown in the limit where
$\widetilde{u}_R$, $\widetilde{d}_R$ and $\widetilde{Q}_L$ are decoupled respectively, and the
coupling constants ($\lambda_i$) are assumed to be equal to unity.  The maximum
positive value is found to be greater than 0.05 for $\Delta \sigma / \sigma$, which is larger
than the limiting case, but this is due to the running effects which induce a small
isospin violating effect on otherwise non-isospin violating simplified models~\cite{DeramoEtAl}.
Thus, if it was found that there was no relative difference between Xe and Ge DM cross sections, 
this could potentially indicate some underlying isospin violation and the existence of
coupling to left-handed (LH) and right-handed (RH) quarks, as the slight miss-match in scalar masses 
is actually required to cancel the isospin violating effects found in~\cite{DeramoEtAl}.  Fig.~\ref{DDExact} shows
the case for finite fixed values of $\Lambda_{Q_L}$, and $\lambda_i$'s are again set to unity for
simplicity.  Note the 
generally non-linear relationship between $\Lambda_{u_R}$
and $\Lambda_{d_R}$ that is required to entirely eliminate isospin violating effects in direct
detection.

\section{Collider Signatures}
The direct detection experiments probe the mediators in combination of 
$\lambda^2/m_{\rm mediator}^2$, with some additional $\lambda$ dependence
from the running effects~\cite{DeramoEtAl}, whereas colliders can probe $m_{\rm mediator}$ 
directly if the mediators can be pair-produced and decay into dijet + \ET.   
If the scalars cannot be pair-produced, then the mono-X + \ET signatures (X = g, W,Z, etc.) will constrain
$\lambda^2/m_{\rm mediator}^2$ as direct detection experiments do.
In the following, we will see that mono-$W$+ \ET is unique in that it probes only 
one mass scale $\widetilde{Q}_L$, because of the LH interactions of $W$ with the SM 
fermions. Other cases always involve both the LH and the RH fermions and their dark scalar partners,
and thus depend on at least two independent mass scales.  Therefore it is important to study 
mono-$W$ + \ET, since it can separately probe the LH sector only.

%The mono-$X$ + missing $E_T$ signatures at hadron 
%colliders, where $X = g, W, Z,$ etc.. 

\subsection{Mono-$W$ + missing $E_T$}
\begin{figure}[htb]
\begin{minipage}{0.5\textwidth}
\centering
\includegraphics[scale=0.8,bb = 175bp 510bp 500bp 795bp,clip]{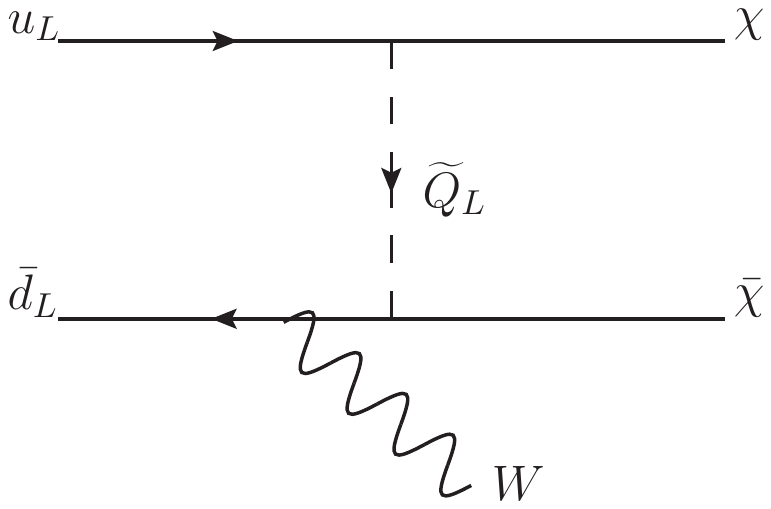}
\end{minipage}
\begin{minipage}{0.5\textwidth}
\centering
\includegraphics[scale=0.8,bb = 175bp 442bp 500bp 735bp,clip]{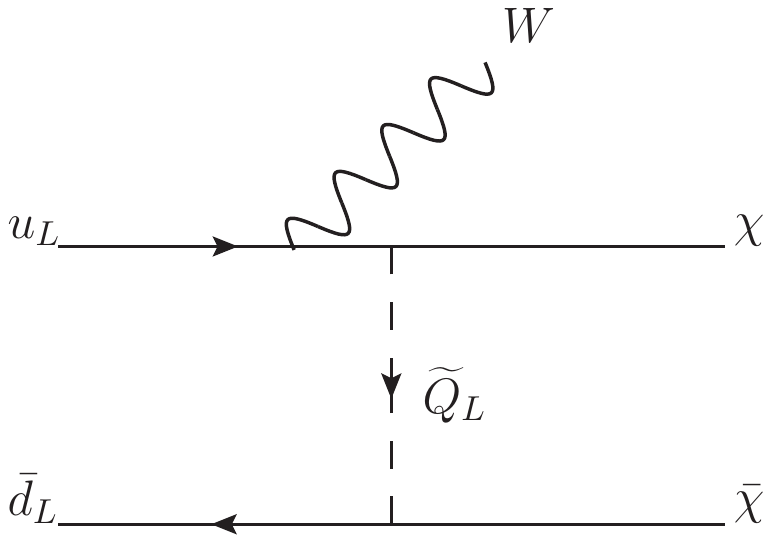}
\end{minipage}
\caption{Feynman diagrams for mono-$W$ that produce $\mathcal{O}(6)$
EFT operators.}
\label{FeynMonoW}
\end{figure}
As stated earlier, the mono-$W$
 mode is nice in that it probes the LH quark sector only, since only $\widetilde{Q}_{L1} 
\equiv ( \widetilde{u}_L , \widetilde{d}_L )$, $\widetilde{Q}_{L2} \equiv ( \widetilde{c}_L , \widetilde{s}_L )$ 
and $\widetilde{Q}_{L3} \equiv ( \widetilde{t}_L , \widetilde{b}_L )$ contribute (the top contribution would be 
negligible for the 13 TeV LHC).   At parton level, there are three Feynman diagrams that 
contribute to the processes 
\(
u \overline{d} \rightarrow W^+ + \chi \overline{\chi}  
\), however as pointed out in the literature the internal Bremsstrahlung 
occurs at a higher order in an EFT~\cite{BellEtAl2015}, 
and large isospin violating effects between $\widetilde{Q}_{Lu}$ and $\widetilde{Q}_{Ld}$
due to $\lambda_4 \neq 0$ provide only a very small 
enhancement of the mono-$W$ relative to mono-jet~\cite{BellEtAl2016}, so for the
purposes of this paper only the processes depicted in Fig.~\ref{FeynMonoW} were
analyzed after verifying the internal Bremsstrahlung and isospin violation, due 
to the $\lambda_4$ mass-splitting, were small.

In our simplified model, the parton level amplitude is given by 
\begin{equation}
\label{parton-monoW}
{\cal M} = ({\cal M}^{\mu}_a + {\cal M}^{\mu}_{b} + {\cal M}^{\mu}_{c})\epsilon^{*}_{\mu}(q),
\end{equation}
where
\begin{equation}
 {\cal M}_a^{\mu} = \frac{g_w \lambda_{Q_L}^2}{\sqrt{2}} \frac{\overline{v}(p_2) \gamma^{\mu} P_L (\slashed{q} - \slashed{p}_2) v(k_2) \overline{u}(k_1) P_L u(p_1) }{\left( p_2 -q \right)^2 \left( \left( p_1-k_1 \right)^2-m^2_{\widetilde{Q}_L}\right)}, 
\end{equation} 
\begin{equation}
{\cal M}_{b}^{\mu}= \frac{g_w \lambda_{Q_L}^2}{\sqrt{2}} \frac{\overline{v}(p_2) P_R v(k_2) \overline{u}(k_1) P_L (\slashed{p}_1 - \slashed{q}) \gamma^{\mu} u(p_1) }{\left( p_1 -q \right)^2 \left( \left( p_2-k_2 \right)^2-m^2_{\widetilde{Q}_L}\right)},
\end{equation}
and
\begin{equation}
{\cal M}_{c}^{\mu}= \frac{g_w \lambda_{Q_L}^2}{\sqrt{2}} \frac{\overline{v}(p_2)P_R v(k_2) \overline{u}(k_1) P_L u(p_1)}{\left( \left( p_1-k_1 \right)^2-m^2_{\widetilde{Q}_L}\right) \left( \left( p_2-k_2 \right)^2-m^2_{\widetilde{Q}_L}\right)} (2k_1 - 2 p_1-q)^{\mu}
\end{equation}
After suitable Fierz transformation~~\cite{BelletAl2011,Nishi} these amplitudes can be written as:
\begin{equation}
{\cal M}_{a}^{\mu} = \frac{g_w \lambda_{Q_L}^2}{2 \sqrt{2}} \frac{\overline{v}(p_2) \gamma^{\mu} P_L (\slashed{q}-\slashed{p}_2)\gamma_{\alpha} u(p_1)\overline{u}(k_1)P_L \gamma^{\alpha} v(k_2)}{\left( p_2 -q \right)^2 \left( \left( p_1-k_1 \right)^2-m^2_{\widetilde{Q}_L}\right)},
\end{equation}
\begin{equation}
{\cal M}_{b}^{\mu} = \frac{g_w \lambda_{Q_L}^2}{2 \sqrt{2}} \frac{\overline{v}(p_2) \gamma_{\alpha} P_L (\slashed{p}_1-\slashed{q})\gamma^{\mu} u(p_1)\overline{u}(k_1)P_L \gamma^{\alpha} v(k_2)}{\left( p_1 -q \right)^2 \left( \left( p_2-k_2 \right)^2-m^2_{\widetilde{Q}_L}\right)},
\end{equation}
and
\begin{equation}
{\cal M}_{c}^{\mu} = \frac{g_w \lambda_{Q_L}^2}{2 \sqrt{2}} \frac{\overline{v}(p_2)\gamma_{\alpha} P_L u(p_1) \overline{u}(k_1) P_L \gamma^{\alpha} v(k_2)}{\left( \left( p_1-k_1 \right)^2-m^2_{\widetilde{Q}_L}\right) \left( \left( p_2-k_2 \right)^2-m^2_{\widetilde{Q}_L}\right)} (2k_1 - 2 p_1-q)^{\mu}
\end{equation}
In the limit $m_{\widetilde{Q}_L} \rightarrow \infty$, the above amplitude is simplified as 
\begin{equation}
{\cal M}_{\rm EFT} =  {\cal M}_{\rm EFT}^{\mu} \epsilon^{*}_{\mu}(q),
\end{equation}
where
\begin{eqnarray}
{\cal M}_{\rm EFT}^{\mu} & = &  \frac{-g_w \lambda_{Q_L}^2}{2 \sqrt{2} m^2_{\widetilde{Q}_L}}   \left( \frac{\overline{v}(p_2) \gamma^{\mu} P_L (\slashed{q}-\slashed{p}_2)\gamma_{\alpha} u(p_1)\overline{u}(k_1)P_L \gamma^{\alpha} v(k_2)}{\left( p_2 -q \right)^2} \right.
\nonumber \\
& + & \left. \frac{\overline{v}(p_2) \gamma_{\alpha} P_L (\slashed{p}_1-\slashed{q})\gamma^{\mu} u(p_1)\overline{u}(k_1)P_L \gamma^{\alpha} v(k_2)}{\left( p_1 -q \right)^2} \right)
\end{eqnarray}
which can be derived from the effective Lagrangian from Eq.~\ref{DD-Fierz} only the LH quark terms.  
Hence, this process depends only one mass scale $m_{\widetilde{Q}_L}$ and one Yukawa coupling $\lambda_{\widetilde{Q}_L}$.

\subsection{Mono-jet + missing $E_T$}

\begin{figure}[htb]
\begin{minipage}{0.5\textwidth}
\centering
\includegraphics[scale=0.8,bb = 175bp 450bp 580bp 795bp,clip]{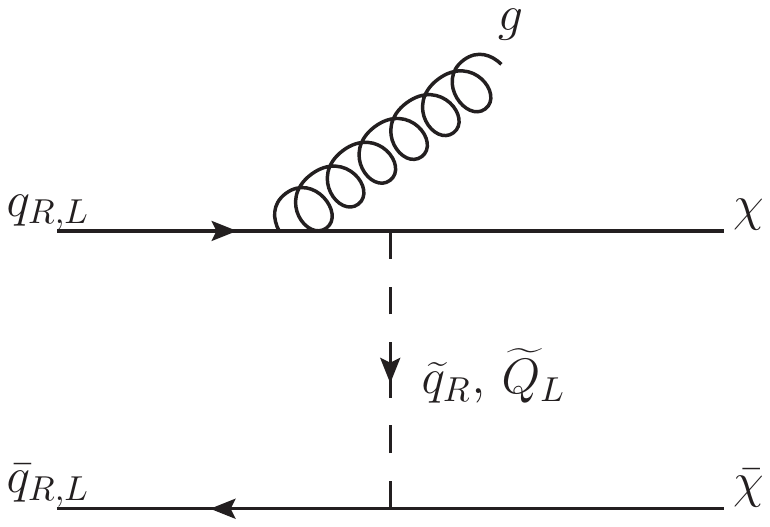}
\end{minipage}
\begin{minipage}{0.5\textwidth}
\centering
\includegraphics[scale=0.8,bb = 175bp 550bp 480bp 800bp,clip]{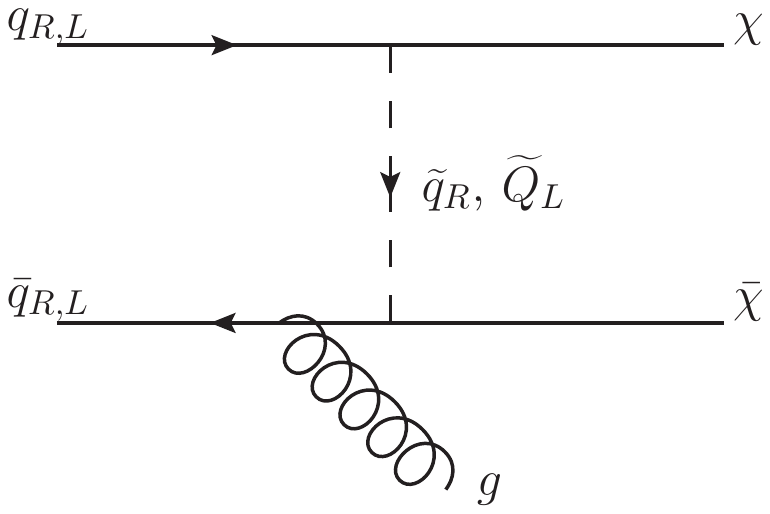}
\end{minipage}
\caption{Feynman diagrams for mono gluon that produce $\mathcal{O}(6)$
EFT operators.}
\label{FeynMonoG}
\end{figure}

\begin{figure}[htb]
\centering
\begin{minipage}{0.45\textwidth}
\centering
\includegraphics[width=\textwidth,bb = 150bp 570bp 415bp 720bp,clip]{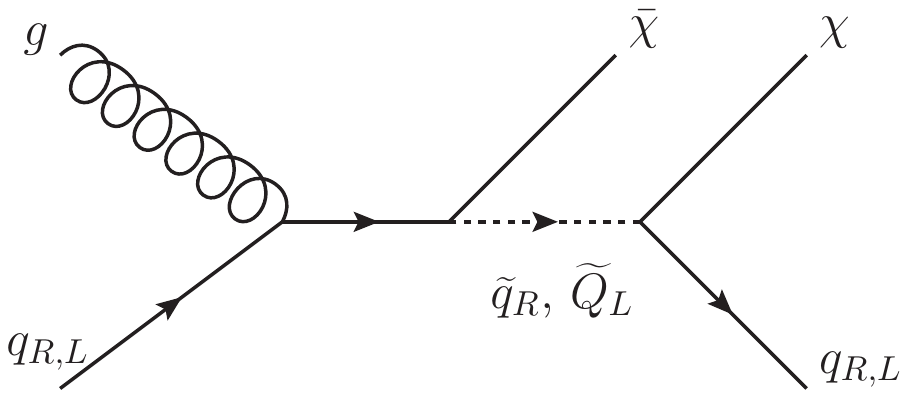}
\end{minipage}
\hfill
\begin{minipage}{0.45\textwidth}
\includegraphics[width=\textwidth,bb =150bp 570bp 415bp 740bp,clip]{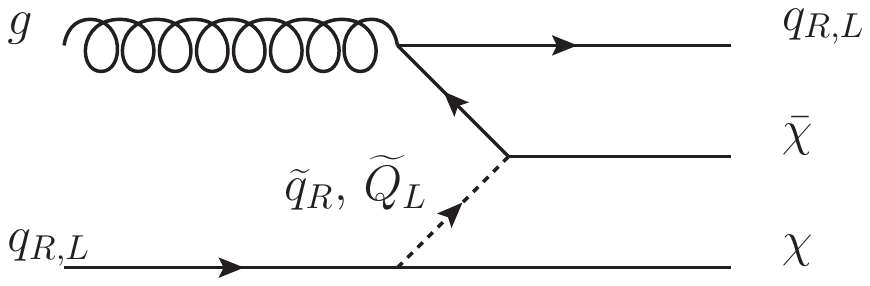}
\end{minipage}
\caption{Feynman diagrams for g q $\rightarrow$ q $\chi \chi$ process at
$\mathcal{O}(6)$ in an EFT.}
\label{FeynMonoJ}
\end{figure}

Since it is very difficult determining a quark-jet from a gluon-jet, the
mono-jet channel includes both mono-$g$ + \ET and mono-$q$ + \ET channels.  
The Feynman diagrams that contribute to the lowest order
in the EFT theory for each separate channel are illustrated in Fig.~\ref{FeynMonoG} and
Fig.~\ref{FeynMonoJ}.

In this channel both the LH and the RH quarks contribute with equal weights.  

The parton level amplitude for  
\[
q \bar{q} \rightarrow g + \chi \bar{\chi} 
\] 
(with $q=u,d,s,c,b$)  is given by 
\begin{equation}
\label{parton-monoG}
{\cal M} = ({\cal M}_{a,L}^{\mu}+{\cal M}_{b,L}^{\mu}+{\cal M}_{c,L}^{\mu}+{\cal M}_{a,R}^{\mu}+{\cal M}_{b,r}^{\mu}+{\cal M}_{c,R}^{\mu})\epsilon_{\mu}^{*}(q),
\end{equation}
\begin{eqnarray}
{\cal M}_{a,L}^{\mu} &=& \frac{g_s \lambda_{Q_L}^2}{2} \frac{\overline{v}(p_2)\gamma^{\mu} P_L \slashed{q}_a \gamma_{\alpha} u(p_1) \overline{u}(k_1) P_L \gamma^{\alpha}v(k_2)}{\left( p_2 -q \right)^2 \left( \left( p_1-k_1 \right)^2-m^2_{\widetilde{Q}_L}\right)}\\
{\cal M}_{b,L}^{\mu} &=& \frac{g_s \lambda_{Q_L}^2}{2} \frac{\overline{v}(p_2)\gamma_{\alpha} P_L \slashed{q}_b \gamma^{\mu} u(p_1) \overline{u}(k_1) P_L \gamma^{\alpha}v(k_2)}{\left( p_1 -q \right)^2 \left( \left( p_2-k_2 \right)^2-m^2_{\widetilde{Q}_L}\right)}\\
{\cal M}_{c,L}^{\mu} &=& \frac{g_s \lambda_{Q_L}^2}{2} \frac{\overline{v}(p_2)\gamma_{\alpha} P_L u(p_1) \overline{u}(k_1) P_L \gamma^{\alpha} v(k_2)}{\left( \left( p_1-k_1 \right)^2-m^2_{\widetilde{Q}_L}\right) \left( \left( p_2-k_2 \right)^2-m^2_{\widetilde{Q}_L}\right)} (2k_1 - 2 p_1-q)^{\mu}\\
{\cal M}_{a,R}^{\mu} &=& \frac{g_s \lambda_{q_R}^2}{2} \frac{\overline{v}(p_2)\gamma^{\mu} P_R \slashed{q}_a \gamma_{\alpha} u(p_1) \overline{u}(k_1) P_R \gamma^{\alpha}v(k_2)}{\left( p_2 -q \right)^2 \left( \left( p_1-k_1 \right)^2-m^2_{\widetilde{q_R}}\right)}\\
{\cal M}_{b,R}^{\mu} &=& \frac{g_s \lambda_{q_R}^2}{2} \frac{\overline{v}(p_2)\gamma_{\alpha} P_R \slashed{q}_b \gamma^{\mu} u(p_1) \overline{u}(k_1) P_R \gamma^{\alpha}v(k_2)}{\left( p_1 -q \right)^2 \left( \left( p_2-k_2 \right)^2-m^2_{\widetilde{q_R}}\right)}\\
{\cal M}_{c,R}^{\mu} &=& \frac{g_s \lambda_{q_R}^2}{2} \frac{\overline{v}(p_2)\gamma_{\alpha} P_R u(p_1) \overline{u}(k_1) P_R \gamma^{\alpha} v(k_2)}{\left( \left( p_1-k_1 \right)^2-m^2_{\widetilde{q_R}}\right) \left( \left( p_2-k_2 \right)^2-m^2_{\widetilde{q_R}}\right)}(2k_1 - 2 p_1-q)^{\mu}
\end{eqnarray}
In the limit $m_{\widetilde{Q}_L} \rightarrow \infty$, the above amplitude is simplifed as 
\begin{eqnarray}
& {\cal M}_{\rm EFT} & = \frac{- g_s}{2} \left[ \frac{\lambda_{Q_L}^2}{ m^2_{\widetilde{Q}_L}} \left( \frac{\overline{v}(p_2) \gamma^{\mu} P_L \slashed{q}_a \gamma_{\alpha} u(p_1) \overline{u}(k_1) P_L \gamma^{\alpha} v(k_2)}{\left( p_2-q \right)^2} 
  +   \frac{\overline{v}(p_2) \gamma_{\alpha} P_L \slashed{q}_b \gamma^{\mu} u (p_1) \overline{u}(k_1) P_L \gamma^{\alpha} v(k_2)}{\left( p_1-q \right)^2}\right) \right. \nonumber \\
 & + & \left. \frac{\lambda_{q_R}^2}{ m^2_{\widetilde{q_R}}} \left( \frac{\overline{v}(p_2) \gamma^{\mu} P_R \slashed{q}_a \gamma_{\alpha} u(p_1) \overline{u}(k_1) P_R \gamma^{\alpha} v(k_2)}{\left( p_2-q \right)^2} 
  +  \frac{\overline{v}(p_2) \gamma_{\alpha} P_R \slashed{q}_b \gamma^{\mu} u (p_1) \overline{u}(k_1) P_L \gamma^{\alpha} v(k_2)}{\left( p_1-q \right)^2}\right)\right]
\end{eqnarray}
which can be derived from the effective Lagrangian in Eq.~\ref{DD-Fierz}.
%The parton level cross section in the full theory is given by
%\begin{equation}
%\hat{\sigma} =  
%\end{equation}
%whereas the one in the EFT is given by 
%\begin{equation}
%\hat{\sigma}_{\rm EFT} =  
%\end{equation}

%\subsection{Mono-q + missing $E_T$}

The mono-quark channel Feynman diagrams are shown in Fig.~\ref{FeynMonoJ}, where
higher order EFT terms are neglected.  In this channel,
the width of the mediator potentially becomes important, and has been included
when calculating the full mono-jet cross section.  While higher multiplicity jet events
are included in LHC jet searches, these were not calculated for this paper.  
Because both the mono-gluon and mono-quark channels depend on all three mediators,
and their couplings, certain assumptions must be made in order to make
LHC predictions.  

%\subsection{Mono-Z + missing \ET}
%In this channel both the LH and the RH quarks contribute with different
%weights determined by the $SU(2)_L \times U(1)_Y$ charges of $u_{L,R}$ and $d_{L,R}$.
%As discussed in previous sections, understanding which operators
%exist at higher energy is relevant for direct dection, and so the mono-X collider searches
%and direct detection
%have a complementarity in determining the structure of the DM mediators.  

\section{Collider Results \& Discussion}

In this section, we present collider phenomenology of mono-$X$+ missing \ET based on the simplified models
with SM gauge symmetry, where $X$ = $W/Z$, or jet.  Contributions to the mono-$Z$ + \ET signature includes both the LH and RH
quarks, which contribute with different weights as determined by the $SU(2)_L \times U(1)_Y$ charges
of $u_{L,R}$ and $d_{L,R}$, thus it does not probe a single scale and 
has been included in the collider signature analysis
as it may provide complementary information to the mono-$W$ and mono-jet 
signatures.  In order to perform the collider signature analysis our model is implemented in Feynrules~\cite{feynrules}, where the hadronic level cross section
is calculated in Madgraph 5~\cite{madgraph5} utilizing the NNPDF23 parton distribution function
set~~\cite{nnpdf23}.  Kinematic plots are then produced through the Pythia and
Delphes interfaces~\cite{madgraph5,pythia6,delphes} and analyzed in Root 
~\cite{root}. 
\begin{figure}[htb]
\begin{minipage}{0.8\textwidth}
\centering
\includegraphics[width=\textwidth]{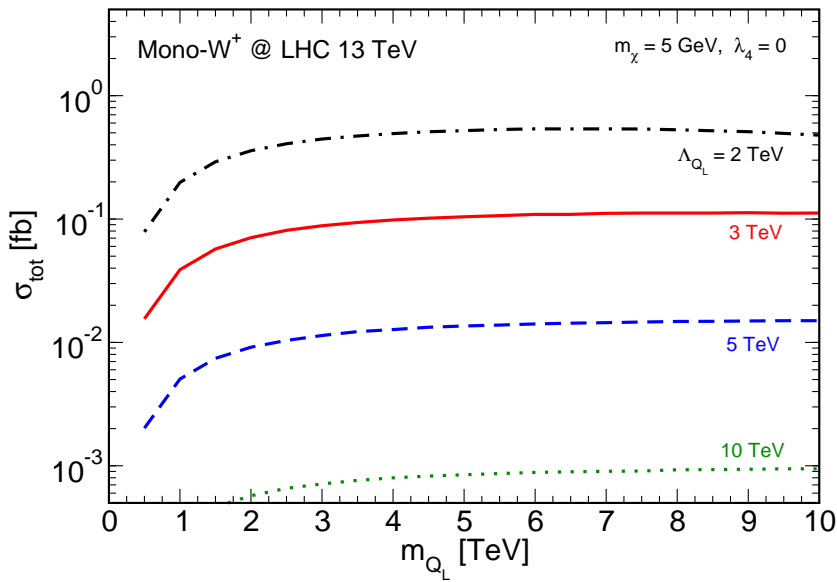}
\end{minipage}
\centering
\caption{Hadron level cross section for pp\(\rightarrow W\) + \ET at 13 TeV.}
\label{CSmonoW}
\end{figure}
\begin{figure}[htb]
\begin{minipage}{0.8\textwidth}
\centering
\includegraphics[width=\textwidth]{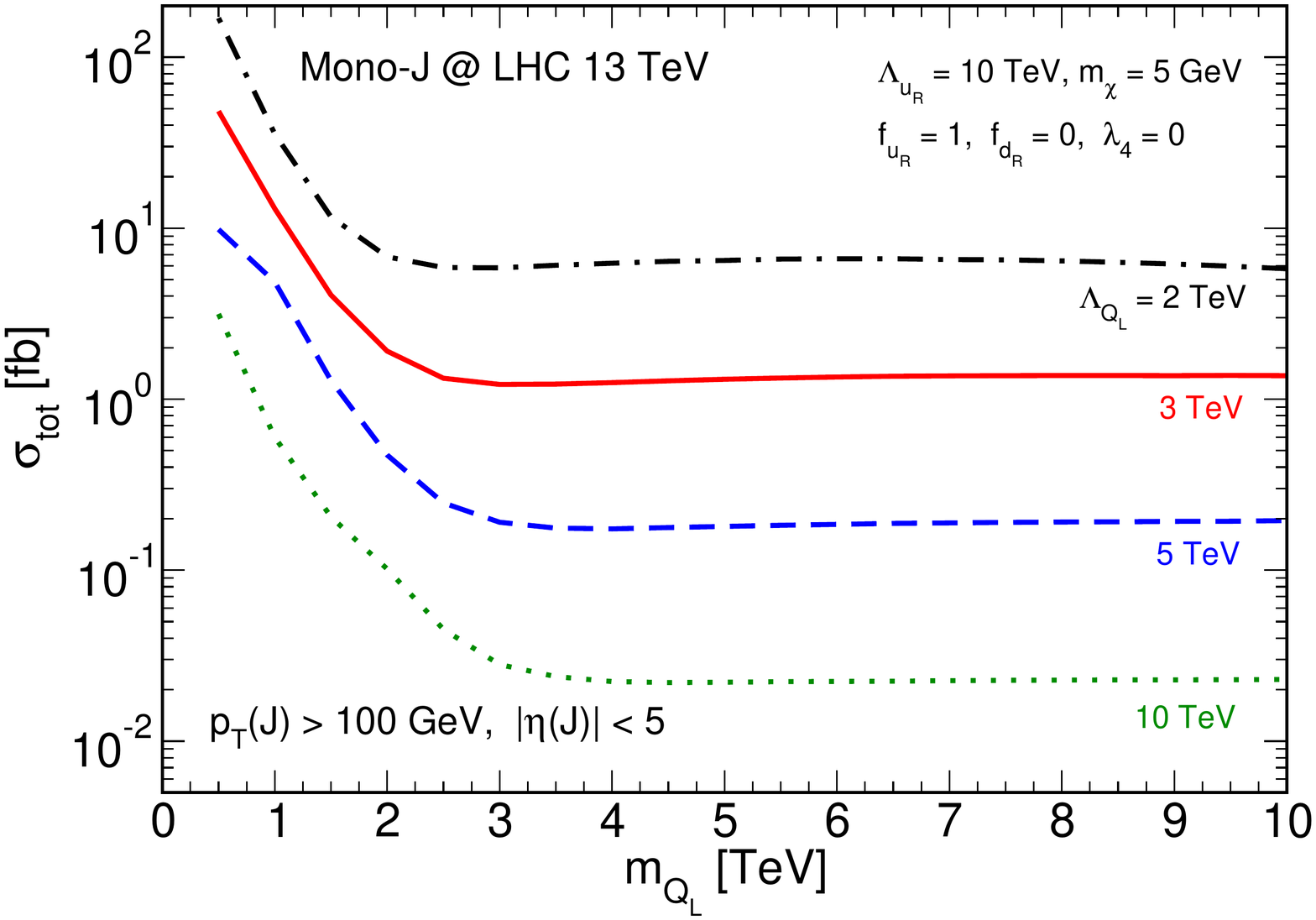}
\end{minipage}
\centering
\caption{Hadron level cross section for pp\(\rightarrow J\) + \ET at 13 TeV.}
\label{CSmonoJ}
\end{figure}
\begin{figure}[htb]
\begin{minipage}{0.8\textwidth}
\centering
\includegraphics[width=\textwidth]{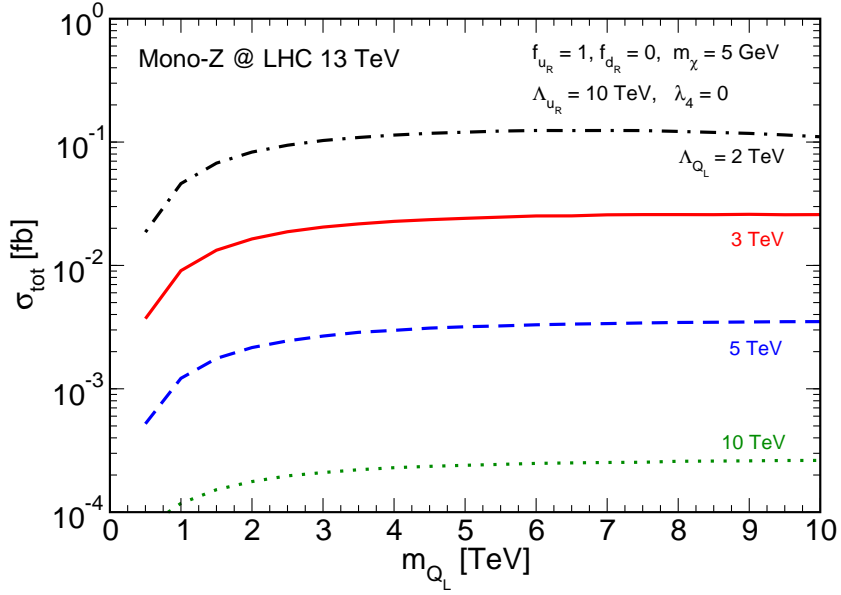}
\end{minipage}
\centering
\caption{Hadron level cross section for pp\(\rightarrow Z\) + \ET at 13 TeV.}
\label{CSmonoZ}
\end{figure}
For the mono-jet search, a minimum jet $p_T$ of 100 GeV is used with a
pseudo-rapidity cut of $| \eta |<$5,
where both the mono-g and mono-quark channels are included.  The minimum scalar mass
is taken to be 1 TeV to account for the jets + \ET searches at the 13
TeV LHC ~\cite{AtlasSquark13TeV2015}.
The hadronic level cross section for the mono-$W^+$ channel is plotted in
Fig.~\ref{CSmonoW} for varying $m_{Q_L}$ and fixed values of
$\Lambda_{Q_L}$ with $m_\chi=5$ GeV, and
the cross section for the mono-jet channel (including both gluon and quark jets)
is plotted in Fig.~\ref{CSmonoJ} with the assumption that $\lambda_{d_R}=0$, $\Lambda_{u_R}=10$ TeV, $m_{\chi}=5$ GeV, and $\lambda_4 = 0$ for varying $m_{Q_L}$ and fixed values of $\Lambda_{Q_L}$.  We find that the mono-$W$ cross sections
are almost flat, and thus are well described by the 
EFT with the cut-off parameter $\Lambda$ except
when $m_{Q_L}$ is close to the dijet limit, where the 
cross section is lower than the EFT prediction.  This is because
the correspondence of $\Lambda_L \leftrightarrow m_{\widetilde{Q}_L}/f_{L}$ 
is violated in the scattering amplitudes due to the typical virtuality of an order of a few TeV of
the $t$-channel mediator.  We find a similar tendency in the
low mediator mass regions in the mono-$Z$ cross sections, which are
shown in Fig.~\ref{CSmonoZ} with the assumptions made for mono-jet
channel in Fig.~\ref{CSmonoJ}.  On the
other hand, for the mono-jet channel the cross section is enhanced
in small $m_{\widetilde{Q}_L}$ due to the $s$-channel pole.

\begin{figure}[H]
\begin{minipage}{0.75\textwidth}
\centering
\includegraphics[width=\textwidth]{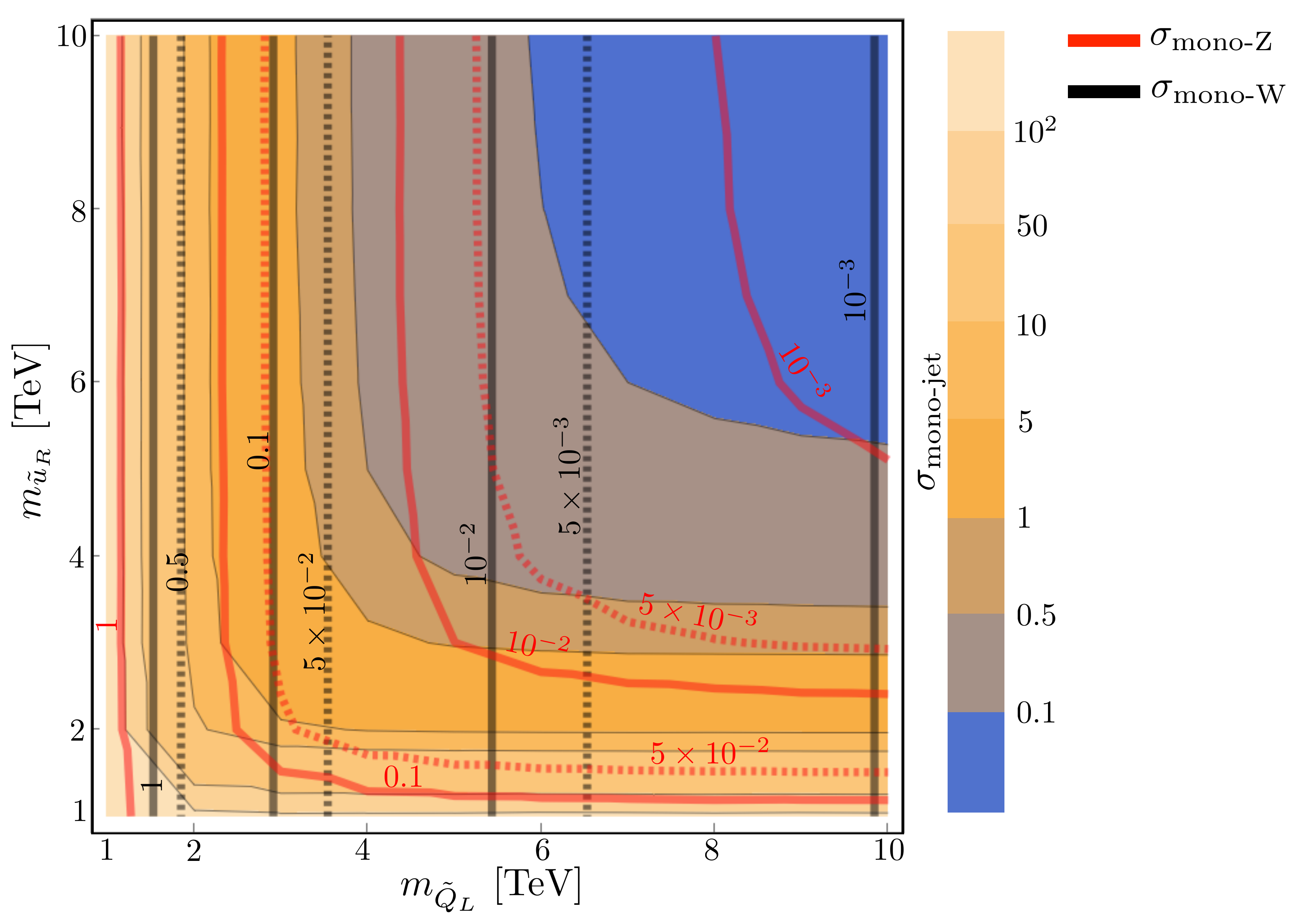}
\end{minipage}
\centering
\caption{Contour plot of $m_{\widetilde{Q}_L}$ vs $m_{\widetilde{u}_R}$
displaying mono-$W$, mono-$Z$, and mono-jet cross sections in fb for
$\lambda_{Q_L}=\lambda_{u_R}=1$, $\lambda_{d_R}=\lambda_4=0$, and $m_{\chi}=5$ GeV.}
\label{uRQL1}
\end{figure}

Note that the mono-$W$ signature is generically
small at the 13 TeV LHC, and so while it provides a weaker constraint on
 $\Lambda$ compared to the mono-jet, it does uniquely provide a constraint
on $\Lambda_{Q_L}$.  The contour plot in Fig.~\ref{uRQL1} shows the
mono-$W$, mono-$Z$ and mono-jet cross sections in fb for $m_{\widetilde{Q}_L}$ versus $m_{\widetilde{u}_R}$ where $\lambda_{Q_L}=\lambda_{u_R}=1$, $\lambda_{d_R}=
\lambda_4=0$, and $m_{\chi}=5$ GeV. 
\begin{figure}[H]
\begin{minipage}{0.75\textwidth}
\centering
\includegraphics[width=\textwidth]{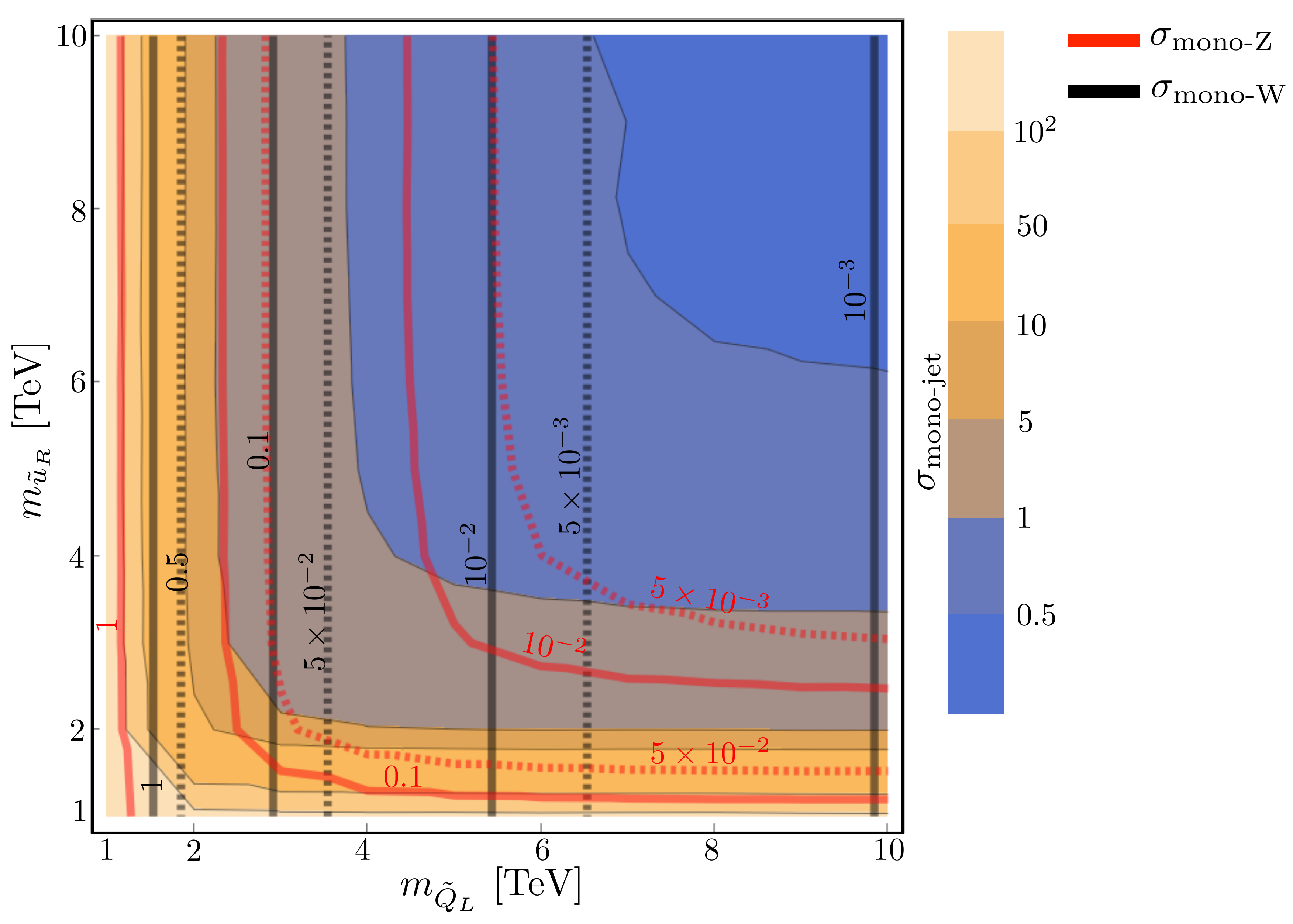}
\end{minipage}
\centering
\caption{The same as Fig.~\ref{uRQL1}, but with
 $\lambda_{\widetilde{d}_R}=1$ and $m_{\widetilde{d}_R}=3$~TeV.}
\label{uRQL2}
\end{figure}
\begin{figure}[H]
\begin{minipage}{0.75\textwidth}
\centering
\includegraphics[width=\textwidth]{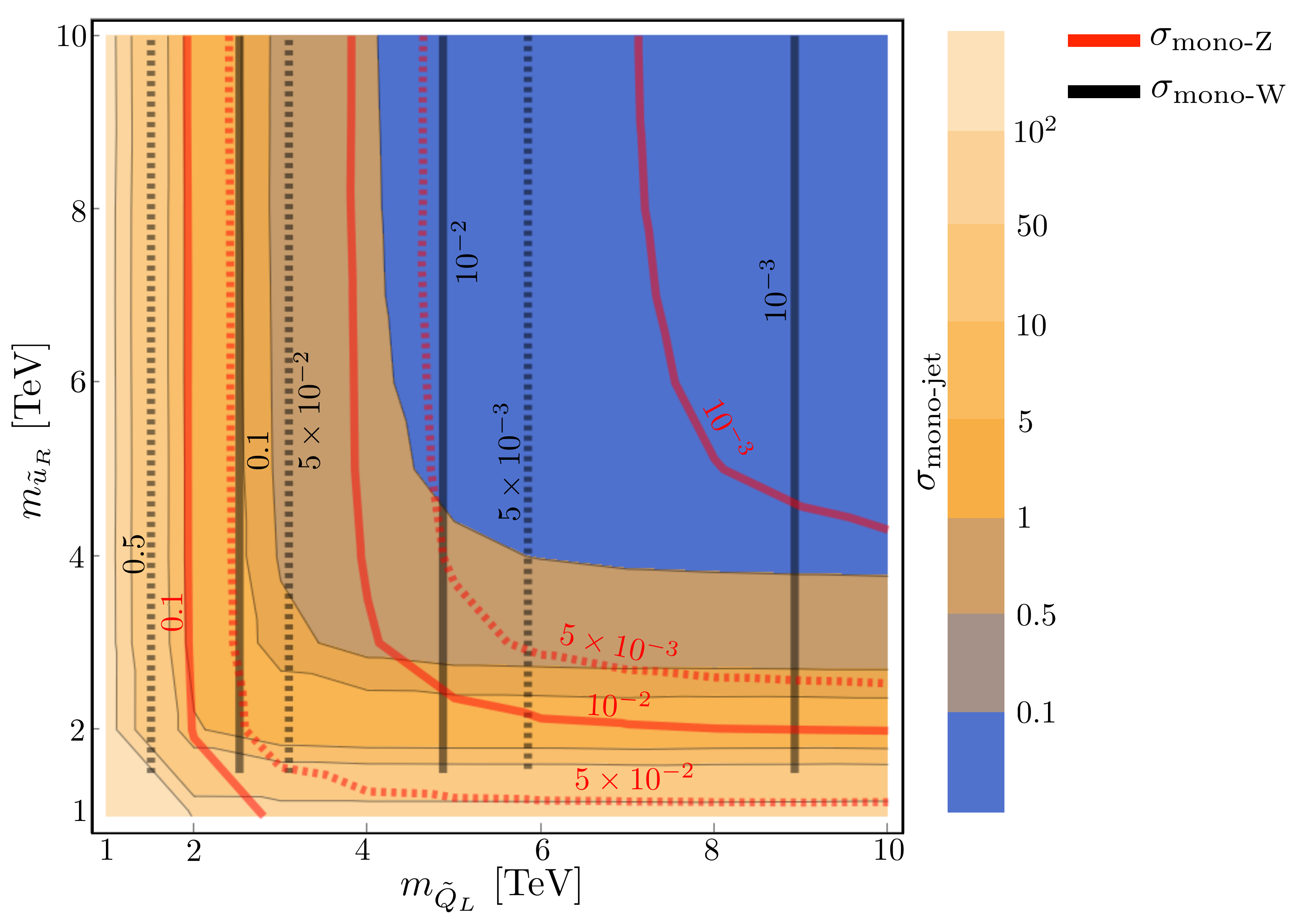}
\end{minipage}
\centering
\caption{The same as Fig.~\ref{uRQL1}, but for $m_{\chi}=300$~GeV.}
\label{uRQL3}
\end{figure}
%
%\begin{figure}[H]
%\centering
%\begin{minipage}{0.45\textwidth}
%\centering
%%\includegraphics[width=\textwidth,bb = 0bp 0bp 540bp 600bp,clip]{/mono-W-J_A}
%\end{minipage}
%\hfill
%\begin{minipage}{0.45\textwidth}
%\centering
%%\includegraphics[width=\textwidth,bb = 0bp 0bp 540bp 600bp,clip]{/mono-W-J_B}
%\end{minipage}
%\centering
%\hfill
%\begin{minipage}{0.45\textwidth}
%\centering
%%\includegraphics[width=\textwidth,bb = 0bp 0bp 540bp 600bp, clip]{/mono-W-J_C}
%\end{minipage}
%\hfill
%\centering
%\begin{minipage}{0.45\textwidth}
%%\includegraphics[width=\textwidth,bb = 0bp 0bp 540bp 600bp,clip]{/mono-W-J_D}
%\end{minipage}
%\centering
%\caption{Mono-jet cross section versus mono-W cross section.}
%\label{csJvscsW}
%\end{figure}
%
The dependence on the mediator masses, and the mono-$X$ cross sections in fb are illustrated
in the contour plot of Fig.~\ref{uRQL1} which shows $m_{\widetilde{Q}_L}$ versus $m_{\widetilde{u}_R}$ for the case when $\lambda_{Q_L}=\lambda_{u_R}=1$, $\lambda_{d_R} = \lambda_4 = 0$ and $m_{\chi}=5$~GeV.  Note that the simplified model case of universal mediator mass and coupling to RH and LH up-quarks is found along the diagonal line starting from the origin, and the case when only one kind of mediator exists is approximated along the
lines when either $m_{\widetilde{Q}_L}$ or $m_{\widetilde{u}_R}$ is 10 TeV.
Previous studies have
shown that failing to account for $SU(2)_L$ gauge invariance has 
lead to an apparent enhancement of the mono-$W$ signature compared to the mono-jet~\cite{BellEtAl2015,BellEtAl2016},
and while our model is $SU(2)_L$ gauge invariant and avoids this problem,
the isospin violation from the LH and RH couplings leads
to certain choices of the parameter space where the mono-jet to mono-$W$ 
ratio is enhanced by nearly three orders of magnitude as seen in the contour plot of 
Fig.~\ref{uRQL1}. Thus, while no large enhancement of the mono-$W$ signature 
can be seen, there is a potentially very large effect in the ratio
of the mono-$W$ and mono-jet signatures which originates in these generic isospin
violating terms and is only visible
 when the free parameters of the simplified model are loosened
from the usual assumptions for the $t$-channel model.  Specifically, take a point along the diagonal where $m_{\widetilde{Q}_L}\approx m_{\widetilde{u}_R}= 2$~TeV and shift this point along a line where one of the $m_{\widetilde{u}_R}=2$~TeV and $m_{\widetilde{Q}_L}>2$~TeV, and the mono-jet goes through one order of magnitude change, mono-$W$ goes through two orders of magnitude, and mono-$Z$ varies by a factor of roughly two.  Thus the parameter space between the case where mediators are treated near universally, compared to the case where all but the $m_{\widetilde{u}_R}$ is too heavy to find at a collider, has a wide range of intermediate predictions that significantly complicate the interpretation of a simplified model constraint derived from LHC data.  Despite
the increased number of free parameters these large deviations in mono-$X$ signature for different assumptions
occurs even when the number of free parameters are restricted such that $\lambda_4=0$,
$\widetilde{u}_R$ is the lightest $t$-channel scalar, $\lambda_{Q_L} \approx 1$, and $\widetilde{d}_R$
can be entirely integrated out.
Fig.~\ref{uRQL2} shows the case where $\widetilde{d}_R$ contribution is
added with fixed $\lambda_{d_{R}}=1$ and $m_{\widetilde{d}_R}=3$~TeV.
For heavier $\widetilde{u}_R$, the mono-jet cross sections is modified,
but the mono-$Z$ cross sections are ostensibly the same as the previous case.  Note that the point where $m_{\widetilde{u}_R}=m_{\widetilde{Q}_L}$ in Fig.~\ref{uRQL2} is equivalent to the simplified model found in Ref.~\cite{DMbenchmark2016} where $m_{mediator}=3$ TeV.  Again, notice that deviations from this point do lead to significant variation in mono-$X$ cross sections.  Importantly, the contour plots of our model allow a practitioner to determine specifically which parameter spaces have overlapping mono-$X$ predictions, to within a factor of 2, simply by looking at which regions overlap.  Fig.~\ref{uRQL3} shows the case where $m_{\chi}=300$~GeV,
while other parameters are the same as those in Fig.~\ref{uRQL1}.  As
discussed in the section on direct detection, the 10 GeV~$< m_{\chi} < 1$ TeV mass range is ruled out for $m_{mediator} \leq 10$~TeV.  However, as described previously, these constraints
can be lifted if we change our assumptions about the nature of $\chi$.  That is, if $\chi$ is a small component of the cosmological DM, or if
it is merely stable for long enough to escape detection at a collider the mass of $\chi$ can be changed.  For example, we find that if $m_{\chi}=300$~GeV and if we assume that there is another species of DM that makes up the majority of the cosmological DM, this mass region is no longer excluded by direct detection.  This different DM mass range has notable effects on the mono-$X$ cross sections as shown in Fig.~\ref{uRQL3}.  

\begin{figure}[H]
\begin{minipage}{0.65\textwidth}
\centering
\includegraphics[width=\textwidth]{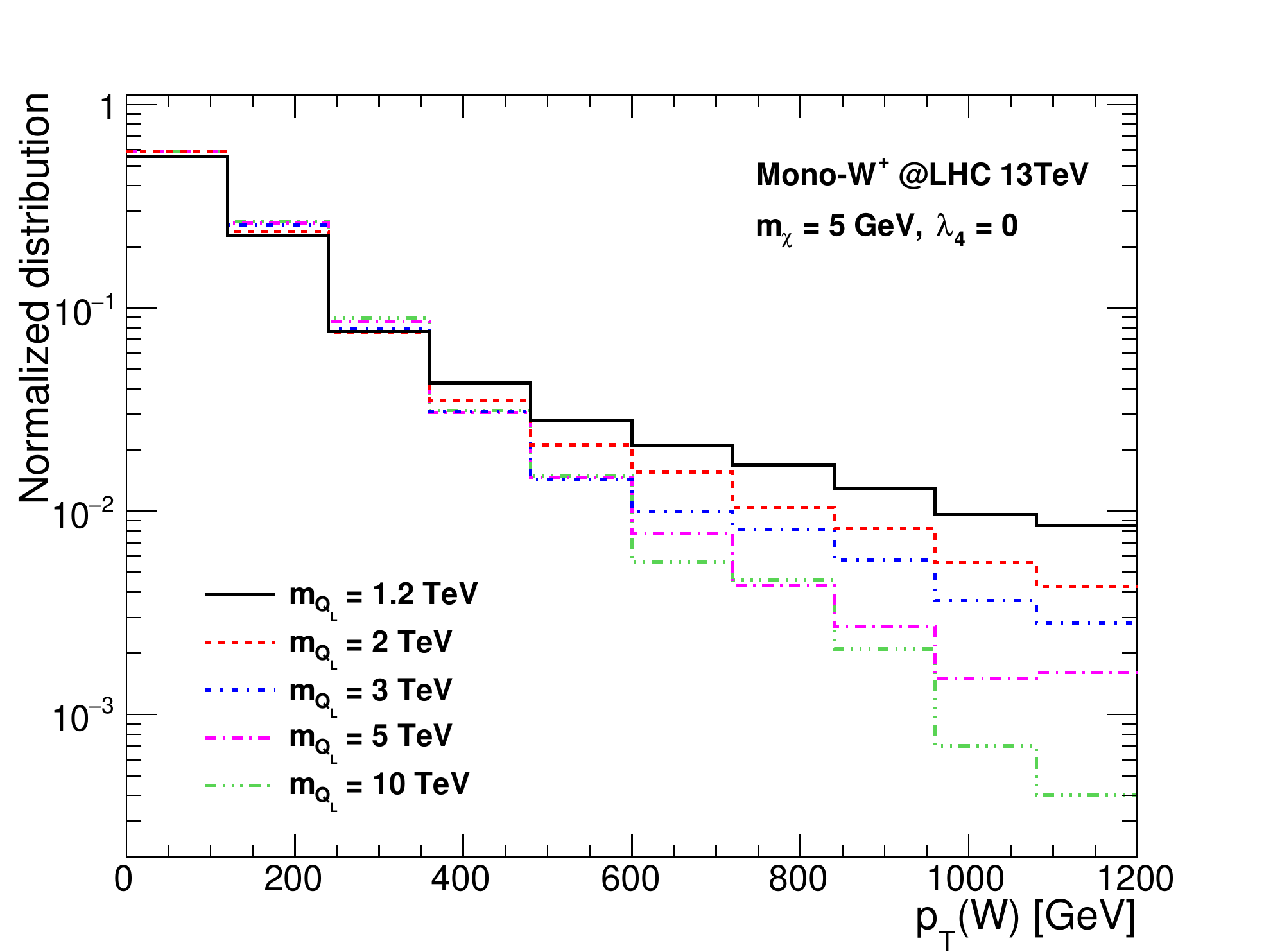}
\end{minipage}
\centering
\caption{Normalized distributions for mono-$W$ $p_T$ when $\lambda_{Q_L}=1$,$\lambda_4=0$, and $m_{\chi}=5$ GeV at parton-level.}
\label{ptdistj}
\end{figure}

\begin{figure}[H]
\begin{minipage}{0.65\textwidth}
\centering
\includegraphics[width=\textwidth]{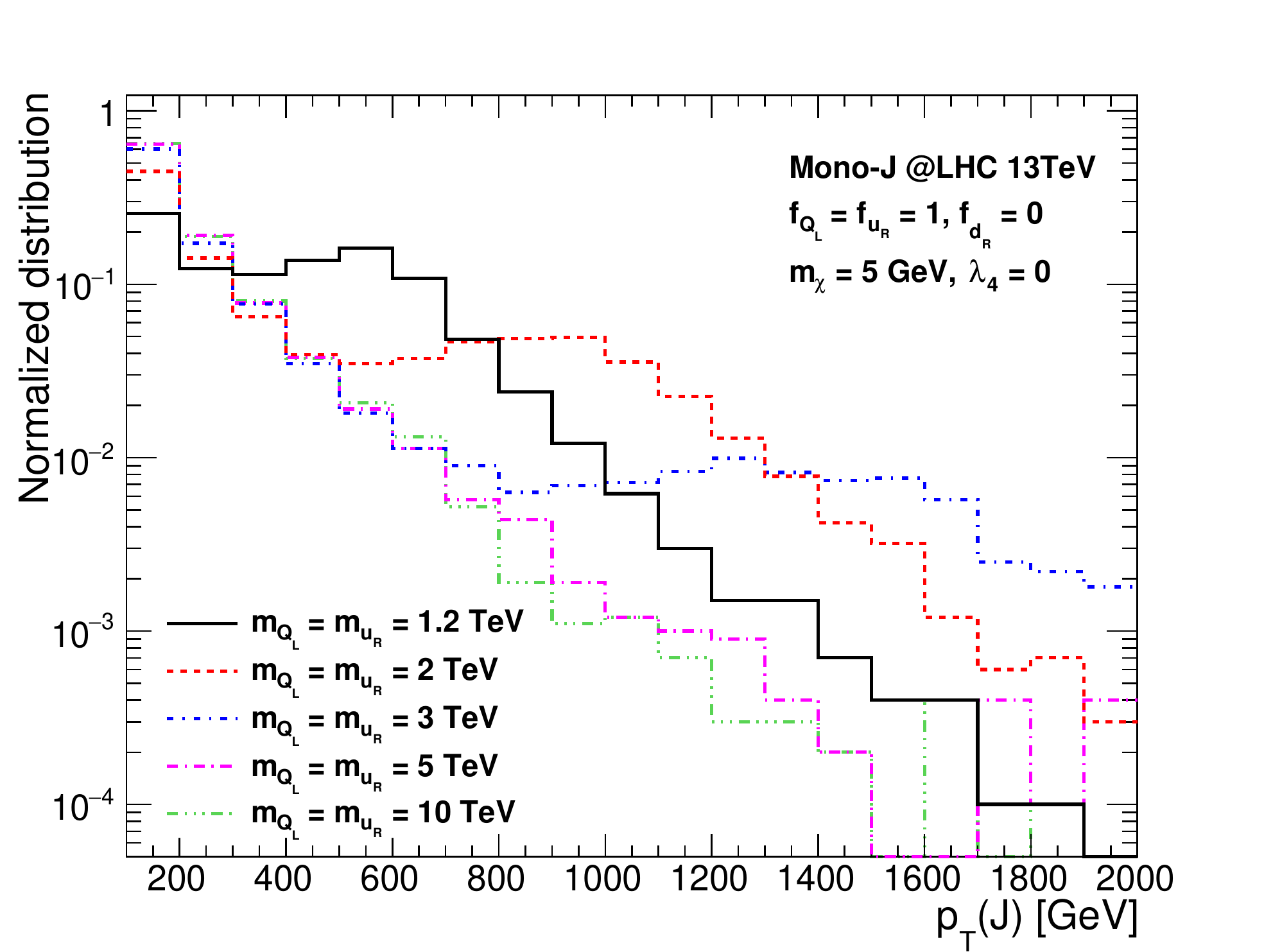}
\end{minipage}
\centering
\caption{Normalized distributions of mono-jet $p_T$ when $\lambda_{Q_L}=\lambda_{u_R}=1$,$\lambda_{d_R}=\lambda_4=0$, and $m_{\chi}=5$ GeV at parton-level.}
\label{ptdistw}
\end{figure}

\begin{figure}[H]
\centering
\begin{minipage}{0.45\textwidth}
\includegraphics[width=\textwidth,bb= 0bp 0bp 185bp 150bp,clip]{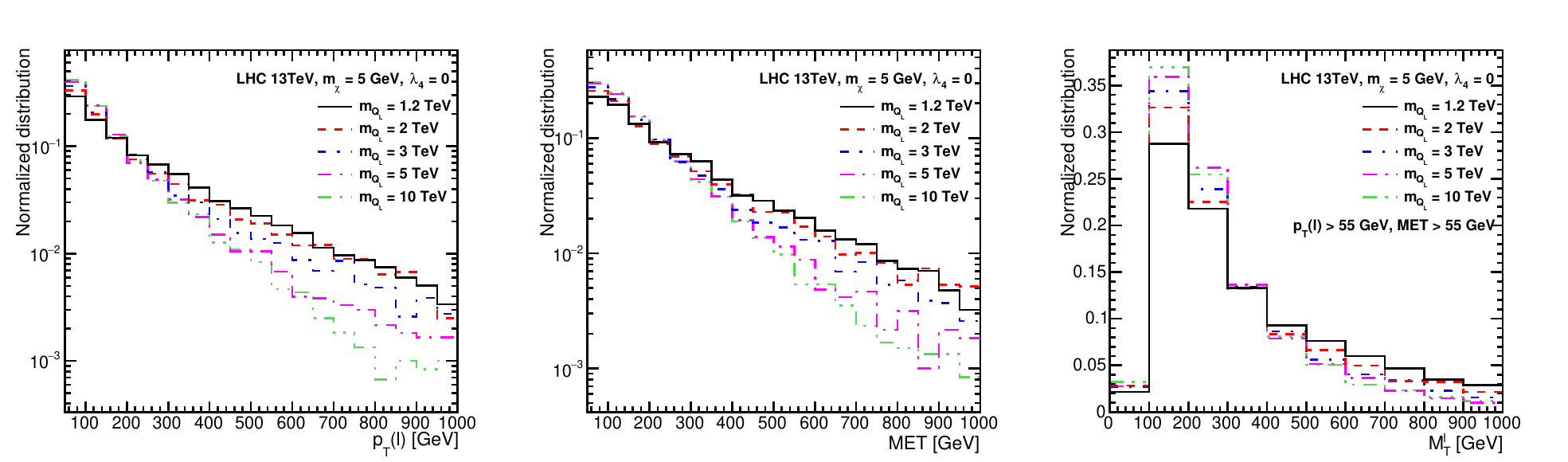}
\end{minipage}
\begin{minipage}{0.45\textwidth}
\includegraphics[width=\textwidth,bb= 185bp 0bp 370bp 150bp,clip]{Graphs/DM5GeV/monoW_Delphes}
\end{minipage}
\hfill
\centering
\begin{minipage}{0.45\textwidth}
\includegraphics[width=\textwidth,bb= 370bp 0bp 555bp 150bp,clip]{Graphs/DM5GeV/monoW_Delphes}
\end{minipage}
\caption{Leptonic observables in mono-$W$ with Delphes simulation for
$\lambda_4=0$ and $m_{\chi}=5$ GeV.}
\label{delphes5}
\end{figure}

Normalized event distributions of $p_T$
for the mono-jet and mono-$W$ signatures are plotted in Fig.~\ref{ptdistj} and Fig.~\ref{ptdistw}.  
For the $p_T$ distribution of the mono-$W$ events, the UV theory has a broader tail in the very high $p_T$ ($>$ 500 
GeV) region, because the small squark virtuality configurations
become relatively important.  On the other hand, the mono-jet 
events have significant peaks centered at around $m_{\widetilde{Q}_L}/2$
due to the Jacobian peak in the scalar decays into a quark and DM.  In Fig.~\ref{delphes5}, we show distributions of letponic observables in mono-$W$ events
after Delphes detector simulation.  The leptonic
observables that are plotted are, moving left-to-right in Fig.~\ref{delphes5}: 
lepton $p_T$, missing $E_T$~(MET), and
the transverse mass $M_T^{\ell}=\sqrt{2p_T^\ell \ET(1-\cos \Delta \phi)}$ where
$\Delta\phi$ is an opening angle of the lepton and missing momentum in the transverse plane.
We find the $p_T$ distribution of the charged lepton and \ET distribution are similar with the $p_T$ distribution of 
$W$.  In addition, $M_T^\ell$ distributions are also different for different scalar mediator masses.
Thus, these leptonic observables can be used to determine the internal parameters of the UV-complete
theory, if enough events are collected at the LHC Run-II.

\section{Summary} % \& Discussion}
The typical approach to investigating DM at colliders, particularly the colored $t$-channel scalars, 
is to make basic assumptions about the particle content and the coupling.  Namely, doublet-only, 
up-like signlet only, down-like singlet only, or universal coupling and mass for
all $t$-channel scalars.  Direct detection strongly constrains the mass of dark matter such that any mass
within the range of $10 \text{ GeV} \le m_{\chi} \le 1000 \text{ GeV}$ for a wide range of
colored scalars is ruled out, and there is a tension between the relic density and direct detection
for simplified models that do not include information about the coupling to leptons.  Additionally, both the relic denisty and direct detection constraints can be changed if there are other species of DM, or if the $\chi$ particle is stable enough to escape a detector if  it is produced at a collider, and thus a $m_{\chi}$ within the range where it is ruled out by direct detection may be observed at the LHC in such a scenario, particularly we have looked at the case when $m_{\chi}=300$~GeV.

For many of the parameter choices presented
in this study the  usual assumptions used for simplified models are justified, however for particular choices of parameters there are striking variations 
in the experimental predictions.  In general,
running effects make it inappropriate to restrict the model to 
a single direct detection operator defined
at high energy~\cite{DeramoEtAl}.  Generically, even if the UV-complete model produced a
single set of operators, such as D5 or D7 as defined in Ref.~\cite{HaiboTait}, 
these running effects would mix the operators making the direct comparison
of LHC Run-I searches to direct detection inappropriate even under the assumption that DM is a single species and $\chi$ is absolutely stable.  Moreover, a UV-complete
model that respects the full SM gauge symmetry, and properly accounts for $SU(2)_L$ 
invariance could have multiple relevant operators, as discussed in this paper. 
These running effects generically generate isospin violating effects
in direct detection, but additional effects can be seen in detectors when $\lambda_{u_R} \neq
\lambda_{d_R}$ as is generally the case
in a UV-complete model and, in fact, $\lambda_{u_R} \neq \lambda_{d_R}$ is required
in our
model to counter-act the generic isospin violating effects seen in Ref.~\cite{DeramoEtAl}, thus a seeming null result in isospin violation at a direct detection experiment may actually imply such isospin violation exists at higher energies.
Of particular note is the large effect these different
assumptions can have on collider signatures, 
where for a given $m_{\widetilde{Q}_L}$ and $\lambda_{Q_L}$
the mono-$W$ signature is the same, but the mono-jet signature can vary by as much as 
three orders of magnitude.  In fact, as parameters are varied from a previously considered simplified model to a parameter space that approximates another simplified model, these mono-$X$ signatures generically change by significant amounts.  From a practitioner's perspective, using a simplified model with a design philosophy similar to ours yields contour plots of the various mono-$X$ cross sections which allow the efficient determination of where in the parameter space simplified models have overlapping or distinct predictions, and thus where an LHC signature could, or could not, uniquely determine the properties of the dark sector.  Thus, while simplified models are still an important tool to understand DM physics at colliders, the broader framework of using the full SM gauge symmetry as discussed above, and allowing for more parameters to vary independently, allows the investigation of parameter space of mono-$X$ signatures that are otherwise ignored.

%{\color{blue} 
%\section*{Note Added}
%While we were completing this work, we received a preprint \cite{D'Eramo:%2016atc} which discusses 
%the importance of the RG evolution when one calculates the DM-nucleon %%scattering cross 
%section in the simplified DM models with the $s$-channel mediators. 
%}

%%% fin

%---------------------------------------------------------------------------
\acknowledgments{
We are grateful to Oliver Buchm\"{u}ller, Patrick Fox, K. Hahn and Lian-Tao Wang, 
for useful discussions on the simplified DM models. 
This work is supported in part by National Research Foundation of 
Korea (NRF) Research Grant NRF-2015R1A2A1A05001869, and by the NRF grant funded 
by the Korea government (MSIP) (No. 2009-0083526) through Korea Neutrino Research Center at Seoul National University (PK), and also by IBS under the project code, IBS-R018-D1 (MP).}%%

%---------------------------------------------------------------------------

%---------------------------------------------------------------------------
%\appendix
%---------------------------------------------------------------------------
%\section{\label{app:same}}
%---------------------------------------------------------------------------

\bibliographystyle{utphys}
\bibliography{Simplified_DM}

%\input{Figures}
%\begin{thebibliography}{999}
%-----------------------------------------------------------------------------
% 1
%\bibitem{CSM}
% \cite{Einhorn:1975ua}
%\bibitem{Einhorn:1975ua}
%  M.B.~Einhorn and S.D.~Ellis,
%  {\it Hadronic production of the new resonances: probing gluon distributions},
%  \prd{12}{1975}{2007};
%  {\it Phys.\ Rev.\ }  {\bf D 12} (1975) 2007.
  %%CITATION = PHRVA,D12,2007;%%
  
%\bibitem{belletal}
%  N.F. ~Bell, J.B. ~Dent, T.D. ~Jacques, and T.J. ~Weiler,
%  {\it W/Z Bremsstrahlung as the Dominant Annhilation Channel for Dark %Matter},
%  {\it Phys.\ Rev.\ } {\bf D 83} (013001) 2011.
  
%\bibitem{nishi}
%  C.C. ~Nishi,
%  {\it Simple derivation of general Fierz-type identities},
%  {\it Am.\ J. \ Phys.} {\bf 73} 1160-1163 2005.
%%-----------------------------------------------------------------------------

%\end{thebibliography}

\end{document}